\documentclass[10pt,a4paper]{article}
\pdfoutput=1
\usepackage{jcappub}
\usepackage{graphicx}
\newcommand{\uvec}[1]{\hat{\boldsymbol{#1}}}
\renewcommand{\vec}[1]{\boldsymbol{#1}}
\usepackage{subcaption}
\usepackage{amsmath,amssymb}
\usepackage{verbatim}
\usepackage{csquotes}
\usepackage{url}

\newcommand{\apjl}{ApJL}
\newcommand{\apjs}{ApJS}
\newcommand{\aap}{A\&A}

\def\araa{\ref@jnl{ARA\&A}}
\usepackage{xcolor}
\definecolor{notecolor}{rgb}{0.8,0,0}
\definecolor{green}{rgb}{0.0, 0.5, 0.0}

\newcommand{\dm}[1]{{\color{magenta} [PDM: #1]}}

\def\lya{Lyman-$\alpha$}
\usepackage{aas_macros}

\title{Revised estimates of CMB $B$-mode polarization induced by patchy reionization}

\author[a,b,c]{Anirban Roy,}
\author[d,e,f]{Girish Kulkarni,}
\author[e,g]{P.~Daniel Meerburg,}
\author[d,e,h]{Anthony Challinor,}
\author[b,c,i,j]{Carlo~Baccigalupi,}
\author[b,c,i,j]{Andrea Lapi,}
\author[d,e]{and Martin~G.~Haehnelt}

\affiliation[a]{Department of Astronomy, Cornell University, Ithaca, NY 14853, USA}
\affiliation[b]{SISSA, Via Bonomea 265, 34136 Trieste, Italy}
\affiliation[c]{Institute for Fundamental Physics of the Universe (IFPU), Via Beirut 2, 34014 Trieste, Italy}
\affiliation[d]{Institute of Astronomy, University of Cambridge, Madingley Road, Cambridge CB3 0HA, UK}
\affiliation[e]{Kavli Institute for Cosmology Cambridge, University of Cambridge, Madingley Road, Cambridge CB3 0HA, UK}
\affiliation[f]{Tata Institute of Fundamental Research, Homi Bhabha Road, Mumbai 400005, India}
\affiliation[g]{Van Swinderen Institute for Particle Physics and Gravity, University of Groningen, Nijenborgh 4, 9747 AG Groningen, The Netherlands}
\affiliation[h]{DAMTP, Centre for Mathematical Sciences, University of Cambridge, Wilberforce Road, Cambridge CB3 0WA, UK}
\affiliation[i]{INFN-Sezione di Trieste, via Valerio 2, 34127 Trieste, Italy}
\affiliation[j]{INAF-Osservatorio Astronomico di Trieste, via Tiepolo 11, 34131 Trieste, Italy}

\emailAdd{ar689@cornell.edu, kulkarni@theory.tifr.res.in, p.d.meerburg@rug.nl, a.d.challinor@ast.cam.ac.uk, carlo.baccigalupi@sissa.it, andrea.lapi@sissa.it, haehnelt@ast.cam.ac.uk}

\abstract{The search for primordial gravitational waves through the
  $B$-mode polarization pattern in the CMB is one of the major goals
  of current and future CMB experiments. Besides foregrounds, a
  potential hurdle in this search is the anisotropic secondary
  $B$-mode polarization generated by the scattering of CMB photons off
  free electrons produced during patchy cosmological
  reionization. Robust predictions of these secondary anisotropies are
  challenging because of uncertainties in the reionization history.
  In this paper we revise estimates of the reionization-induced
  $B$-mode signal by incorporating recent advances in the
  understanding of reionization through observations of the
  Lyman-$\alpha$ forest. To derive these $B$-mode estimates, we use
  high-dynamic-range radiative transfer simulations of reionization
  that are calibrated to the Ly$\alpha$ data.  These simulations are
  also consistent with a variety of other high-redshift observations.
  We find that around multipoles $\ell\approx 100$, reionization
  induces $B$-mode power with $\ell(\ell+1)C_\ell^{BB}/2\pi\approx
  4\times 10^{-6}\,\mu$K$^2$.  This secondary signal is thus at the
  level of the primordial signal with the tensor-to-scalar ratio
  $r<10^{-4}$, and can increase by a factor of $\sim 50$ if
  reionization is sourced by highly clustered sources residing in
  haloes with mass of $\sim 10^{11}$~M$_\odot$.  Our findings suggest
  that the contribution of patchy reionization to the search for
  primordial gravitational waves is unlikely to be a concern for
  currently planned CMB experiments.}


\arxivnumber{---}

\begin{document}

\maketitle
\flushbottom

\section{Introduction}
\label{sec:intro}

An important goal for current and next-generation CMB experiments is the
detection of $B$-mode polarization of the CMB induced by primordial
gravitational waves \cite{2016ARA&A..54..227K}.  These include
the BICEP Array \cite{2018SPIE10708E..07H}, SPT-3G \cite{2014SPIE.9153E..1PB}, Simons
Observatory \cite{2019JCAP...02..056A}, CMB-S4 \cite{2016arXiv161002743A}, LiteBIRD \cite{2014JLTP..176..733M}, and proposed space missions such as PICO
\cite{2018SPIE10698E..4FS}, and CMB-Bharat \cite{CMB-Bharat-Proposal}.
These experiments hope to build on the successes of past experiments,
such as Planck \cite{2018arXiv180706205P}, ACT
\cite{2017JCAP...06..031L}, SPT \cite{2013PhRvL.111n1301H},
BICEP2/Keck \cite{2015ApJ...811..126B} either to detect or
significantly reduce the upper limit on the tensor-to-scalar ratio
parameter $r$, which quantifies the relative power in primordial gravitational waves compared to curvature fluctuations, by one or two orders of magnitude from the current limit
$r<0.06$ (95\,\% confidence) \cite{2018PhRvL.121v1301B}. 
A rather natural target is $r\sim 10^{-3}$ for single-field, slow-roll inflation. 
It has been argued that models that naturally explain the measured tilt of the power spectrum of primordial curvature fluctuations fall into two classes~\cite{Creminelli2015,CMBS42016}: one in which the potential driving inflation is a monomial; and one in which the inflaton field is rolling off a plateau with a potential that varies over some characteristic scale. Models in the former class generally produce significant primordial gravitational waves and current limits on $r$ rule out many examples. In the latter case, $r\lesssim 10^{-3}$ corresponds to models where the characteristic scale is below the Planck scale.
Detection of primordial $B$ modes would significantly strengthen the evidence for inflation while providing a measurement of the energy scale at which inflation occurred and pointers to how inflation is realised in fundamental theory.

In order to succeed, these experiments will have to confront
foreground $B$-mode polarization. The dominant part of these
foregrounds is created by polarized diffuse synchrotron and dust emission 
from the Galaxy \cite{2016A&A...594A...9P},
which can be handled by means of avoidance and/or removal strategies
\cite{2015PhRvL.114j1301B, 2016A&A...594A...9P, 2016A&A...586A.133P}.
Foregrounds with respect to the primary signal from inflation also
consist of secondary $B$-mode polarization anisotropies that can
potentially have a significantly larger amplitude than primordial $B$
modes.  These include secondary $B$ modes induced by gravitational
lensing \cite{1998PhRvD..58b3003Z, 2006PhR...429....1L}, the thermal
and kinetic Sunyaev-Zel'dovich effect \cite{1972CoASP...4..173S,
  1980ARA&A..18..537S}, and by fluctuations in the Thomson scattering
opacity of the Universe during the epoch of reionization
\cite{Gnedin2001, 2005ApJ...630..657Z, Hu2000, Dvorkin2009,
  DvorkinBmode}.  Lensing-induced $B$ modes have now been detected by
several experiments \cite{2013PhRvL.111n1301H, 2014PhRvL.113b1301A,
  2015ApJ...808....7V, 2014ApJ...794..171P, 2014PhRvL.112x1101B,
  2015ApJ...807..151K}.  While this signal can dominate over the
primordial $B$ modes on all but the largest scales for $r\lesssim 10^{-2}$, it is possible to
remove partially the effects of lensing with an estimate of the lensing deflections obtained either from a suitably correlated tracer of large-scale structure or from the small-scale CMB itself~\cite{2002PhRvL..89a1304K, 2002PhRvL..89a1303K, 2004PhRvD..69d3005S}.

Compared to lensing-induced $B$ modes, $B$ modes created due
to patchy reionization are more difficult to disentangle from the
primordial signal. This is because of our current lack of detailed knowledge of how the Universe reionized.
Reionization produces $B$-mode anisotropies due to inhomogeneous scattering of the local quadrupole CMB temperature
anisotropy as well as anisotropic screening of the primary CMB polarization signal sourced at recombination.
Several studies have estimated these $B$-mode signals \cite{2001ApJ...551....3G,
  2005ApJ...630..657Z, 2000ApJ...529...12H, Santos2003, DvorkinBmode,
  Dvorkin2009, Roy2018, 2019arXiv190301994M}.  However, all of these
studies have suffered from our ignorance of when and how
reionization happened.  The average Thomson scattering optical depth inferred from large-angle $E$-mode polarization has changed significantly over the last 15 years, with the central value of the implied redshift of reionization shifting from
$z_\mathrm{re}\sim 20$~\cite{2003ApJS..148....1B} to $z_\mathrm{re}\sim 7$~\cite{Planck2018}.  The best constraints on
the end of reionization have resulted from observations of Lyman-$\alpha$
absorption features in the spectra of bright quasars.
Constraining the earlier stages of reionization has been difficult, however, because
the number of such quasars rapidly decreases at high redshifts.
Additionally, the Lyman-$\alpha$ absorption saturates at neutral
hydrogen fractions of $n_\mathrm{HI}/n_\mathrm{H}\sim 10^{-4}$ due to
the large cross-section of the Lyman-$\alpha$ transition.

A further complication is that the level of reionization-induced
$B$-mode anisotropy also depends on the morphology of the ionized
regions created during reionization.  Reionization is thought to occur
due to the formation and eventual overlap of ionized regions around
sources of hydrogen-ionizing radiation such as galaxies or quasars.
The size and evolution of these ionized regions depend on the mass
and brightness of these sources. However, the detailed nature of these sources remains unclear.
Bright quasars are efficient sources of hydrogen-ionizing
radiation, but their abundance at high redshifts is probably too low
\cite{2018arXiv180709774K}.  Galaxies are known to exist at redshifts
as high as $z\sim 11$ \cite{Bouwens:2015vha}, but it is unclear if hydrogen-ionizing photons
produced by young stars within these galaxies can escape into the
intergalactic medium without being absorbed by the in-situ hydrogen
\cite[e.g.,][]{2018arXiv180303655C, 2018arXiv180511621M,
  2018arXiv180601741F, 2018arXiv180506071S}.  For a given reionization
history, reionization by quasars rather than galaxies can result in a
larger amplitude of the induced secondary $B$ modes due to the
enhanced clustering of quasars \cite{2017MNRAS.469.4283K}. Similar
changes in the $B$-mode anisotropy amplitude can occur due to
different relative contributions of bright and faint galaxies to
reionization. Brighter galaxies typically reside in highly-clustered
massive haloes, which can increase the anisotropy amplitude on certain
scales.

As noted above, it has traditionally been difficult to get tight constraints on the
redshift of reionization because the pre-eminent probe of the
high-redshift intergalactic medium, the Lyman-$\alpha$ forest,
saturates already when the neutral hydrogen fraction is around $10^{-4}$.
In recent years, however, excellent constraints on the
redshift at which reionization ended have come from a somewhat
unexpected direction.  It has been noted that the Lyman-$\alpha$
forest at $z\approx 5.5$ exhibits spatial fluctuations that are
significantly larger than those expected due to density
inhomogeneities in a post-reionization Universe \cite{Fan2006,
  2015MNRAS.447.3402B, 2018MNRAS.tmp.1287B, 2018ApJ...864...53E,
  2018ApJ...863...92B}.  Recently, it was shown that these
fluctuations are a signature of the last stages of reionization
\cite{2019MNRAS.485L..24K}.  In this picture, large patches (of sizes
of up to 100 comoving Mpc) of neutral hydrogen exist in low-density
regions of the Universe at $z\approx 5.5$.  It is these ``neutral
islands'' that cause the observed spatial scatter in the
Lyman-$\alpha$ forest.  This interpretation yields a rather tight
constraint on the redshift of the end of reionization of $z=5.2$. 

Given these developments in our understanding of reionization, providing updated estimates of the reionization-induced
$B$-mode power is timely and of obvious interest to next-generation CMB experiments.
We consider this issue in this paper.

We use hydrodynamic cosmological simulations post-processed for
radiative transfer.  These simulations have a large dynamic range
(from around $100$ comoving kpc to $160$ comoving Mpc) that allows us to
model the effect of small-scale reionization sources while deriving
the large-scale CMB signal.  They are carefully calibrated to
reproduce the Lyman-$\alpha$ forest data, including the spatial
fluctuations in the Lyman-$\alpha$ forest opacity.  In the process,
these simulations also yield a value of the Thomson scattering optical
depth to the CMB last-scattering surface that is in agreement with the
latest measurements reported by Planck 2018 \cite{Planck2018}.  We
extrapolate these simulations to even larger length scales (1 Gpc) by
using the excursion set method of modelling ionized regions in the
epoch of reionization (EoR)\cite{Furlanetto2004}.  After presenting
predictions for the $B$-mode signal in our fiducial reionization
model, we also consider the effect on this signal of the uncertainty
in the first half of the reionization history and the uncertainty regarding the masses of the reionizing sources.  Finally, we consider the range of
values of the tensor-to-scalar ratio $r$ that can be reliably probed
by various upcoming CMB experiments in the presence of the secondary $B$
modes from reionization.

Throughout this work we assume a flat $\Lambda$CDM universe with baryon and matter density parameters $\Omega_\mathrm{b}=0.0482$ and
$\Omega_\mathrm{m}=0.308$,
Hubble constant $100 h\,\mathrm{km}\,\mathrm{s}^{-1}\,\mathrm{Mpc}^{-1}$ with
$h=0.678$, spectral index of primordial curvature perturbations
$n_\mathrm{s}=0.961$, clustering amplitude $\sigma_8=0.829$ at $z=0$, and helium mass fraction $Y_\mathrm{He}=0.24$
\citep{2014A&A...571A..16P}.  The units `ckpc' and `cMpc' refer to
comoving kpc and comoving Mpc, respectively.

\section{Secondary \texorpdfstring{$B$}{Lg}-mode anisotropies from reionization}
\label{patchyBB}

After CMB decoupling at $z \sim 1100$, hydrogen and helium are essentially completely recombined and the gas in the Universe is neutral. Eventually, the first bound objects are expected to appear in the
form of proto-stars around $z \sim 50$ \cite{2006MNRAS.373L..98N}.
By around $z=5$, the ionizing radiation from galaxies and quasars reionizes the Universe~\cite{Barkana2001, 2019MNRAS.485L..24K, Planck2018}. Inhomogeneities in the distribution of matter and ionizing sources mean that
different parts of the Universe will reionize
at different moments in time. One consequence of this
patchy reionization is that the Thomson optical depth $\tau$, as measured to the CMB last-scattering
surface, is a function of position on the sky,
i.e., $\tau(\uvec{n})$.  Patchy reionization produces
secondary anisotropies in the polarization of the CMB through two mechanisms~\cite[e.g.,][]{Hu2000,DvorkinBmode}: (i)
polarization is generated through Thomson scattering of the local CMB temperature quadrupole anisotropy off the inhomogeneously-distributed free electrons during reionization; and (ii) the primary polarization generated by scattering around recombination is screened (i.e., polarized radiation is scattered out of the line of sight) by the anisotropic optical depth $\tau(\uvec{n})$.
While both effects are present
even in the absence of fluctuations in the free
electron density, $B$ modes will not be generated unless there are inhomogeneities.
Patchy reionization
will thus produce a unique signal in the $B$-mode polarization pattern
of the CMB.

The free-electron number density during reionization varies due to inhomogeneities in the gas density and the ionization fraction, $x_e$. Previous semi-analytic work showed that for most plausible models, the power is dominated by the ionization rather than density inhomogeneities~\cite{Mortonson}.
The ionization fraction is defined as $x_e = n_e / n_p$, i.e., the ratio of the number densities of free electrons to hydrogen nuclei. We decompose this into an average fraction $\bar{x}_e$ and a fluctuation $\Delta x_e$, so that at comoving distance $\chi$ along a line of sight $\uvec{n}$ we have 
\begin{equation}
x_e(\uvec{n},\chi)=\bar{x}_e(\chi)+\Delta {x_e}(\uvec{n},\chi),
\end{equation}
where $\bar{x}_e(\chi)$ is the mean fraction at (comoving) lookback time $\chi$ and 
$\Delta {x_e}(\uvec{n},\chi)$ is the fluctuation at position $\uvec{n}\chi$ relative to the observer at lookback time $\chi$.
The optical depth
measures the line-of-sight integral of the electron density, 
and therefore also
becomes a direction-dependent quantity on the sky. The optical depth back to lookback time $\chi$ is approximately
\begin{equation}
\tau(\uvec{n},\chi)=\sigma_{\text{T}} \bar{n}_{p,0}\int_0^\chi
\frac{d\chi'}{a^2}\left[\bar{x}_e(\chi')+\Delta
  x_e(\uvec{n},\chi')\right],
  \label{eq:taudef}
\end{equation}
where
$\sigma_{\text{T}} = 6.652\times 10^{-29}\,\mathrm{m}^2$ is the Thomson
scattering cross section, $\bar{n}_{p,0}$
is the present unperturbed number density of protons and $a$ is the scale
factor.
In the Limber approximation (valid for angular multipoles $\ell
\gtrsim 20$) \cite{PhysRevD.78.123506,1953ApJ...117..134L,
  2017JCAP...05..014L}, the angular power spectrum of the optical depth fluctuations back to CMB decoupling at $\chi_\ast$ is
\begin{equation}
C_\ell^{\tau\tau}=\sigma_{\text{T}}^2 \bar{n}^2_{p0}\int_0^{\chi_\ast} \frac{d\chi}{a^4\chi^2} P_{
  x_e x_e}\left(k=\frac{\ell+1/2}{\chi}; \chi\right).
\label{eq:tautau}
\end{equation}
Here, $P_{x_ex_e}(k;\chi)$ is the dimensional 3D power spectrum of $\Delta x_e$ at lookback time $\chi$,
which satisfies
$\langle
\Delta x_e(\boldsymbol{k},\chi)\Delta x_e(\boldsymbol{k}^\prime,\chi)\rangle=(2\pi)^3\delta^{(3)}(\boldsymbol{k}+\boldsymbol{k}^\prime)P_{x_ex_e}(k;\chi)$. All
the relevant astrophysical aspects of reionization are encoded in
$P_{x_ex_e}(k)$. We note that $C_\ell^{\tau\tau}$ is not directly
observable, although it can be reconstructed from the non-Gaussianity that patchy screening imprints in the small-scale CMB fluctuations~\cite{Roy2018, Dvorkin2009, Su2011} in a similar manner to how the CMB lensing power spectrum is reconstructed.

\subsection{\texorpdfstring{$B$}{Lg}-mode signal due to scattering of the CMB temperature quadrupole}

To calculate the $B$-mode signal from reionization we need to
account for the scattering probability of photons, fluctuations in the
free electron density along the line of sight, and the statistics of the quadrupole moment of the temperature anisotropies through reionization. As the quadrupole of the primary temperature anisotropies, sourced around recombination, has a much
larger correlation length than the fluctuations in the electron density, the $B$-mode power induced by scattering of the primary quadrupole simplifies, in the Limber approximation, to~\cite{Hu2000}
\begin{equation}
C_\ell^{BB\rm{(sca})}=\frac{3}{100}\int
\frac{d\chi}{\chi^2}\, \left[\frac{g(\chi)}{\bar{x}_e(\chi)}\right]^2 Q^2_{\rm rms}(\chi)P_{x_e
  x_e}\left(k=\frac{\ell+1/2}{\chi};\chi\right) .
\label{eq:clbb1}
\end{equation}
Here $g(\chi)$ is the (unperturbed) visibility function, the probability density for a photon to scatter at lookback time $\chi$, which can be written as 
\begin{equation}
g(\chi)=\frac{d\bar{\tau}}{d\chi}e^{-\bar{\tau}(\chi)}=
\frac{\sigma_{\text{T}} \bar{n}_{p,0}}{a^2}\bar{x}_e(\chi)e^{-\bar{\tau}(\chi)}.
\label{eq:g}
\end{equation}
The quantity $Q_{\rm rms}(\chi)$ is the r.m.s.\ of the primary quadrupole at lookback time $\chi$. Given the near scale-invariance of the power spectrum of the primordial fluctuations, $Q_{\rm rms}$ is almost independent of $\chi$ through reionization. We take $Q_{\rm rms} = 17\,\mu{\rm K}$ (e.g.,~\cite{Dvorkin2009, Hu2000}).
Combining the above equations we find
\begin{align}
C_\ell^{BB({\rm
    sca})}=\frac{3\sigma_{\text{T}}^2\bar{n}_{p,0}^2}{100}\int_{\chi_{\rm s}}^{\chi_{\rm e}}\frac{d\chi}{a^4\chi^2}\, Q^2_{\rm
  rms}(\chi) P_{x_e x_e}\left(k=\frac{\ell+1/2}{\chi};\chi\right) e^{-2\bar{\tau}(\chi)},
  \label{eq:BBscatter}
\end{align}
where $\chi_{\rm s}$ and $\chi_{\rm e}$ are the comoving distances to the start and
end of reionization, respectively. Note the similarity to the optical depth power spectrum, Eq.~\eqref{eq:tautau}, showing that the shape of the two spectra are very similar.

In the electron rest-frame, there is a further contribution to the temperature quadrupole anisotropy that is second order in the baryon peculiar velocity (the kinematic quadrupole). $B$-mode polarization is produced by scattering this quadrupole even in the absence of further modulation by the electron number density fluctuations~\cite{Hu2000}. We do not consider this source further in this paper since the modulated signal from patchy reionization is smaller than that from scattering the primordial quadrupole by a factor of $\bar{T}_{\rm CMB}^2 v_{\rm rms}^4/Q_{\rm rms}^2$, where $v_{\rm rms}\sim 10^{-3}/\sqrt{1+z}$ at the relevant redshifts and $\bar{T}_{\rm CMB}$ is the present-day CMB temperature. Furthermore, the kinematic quadrupole has a distinct frequency dependence (e.g., Ref.~\cite{Kamionkowski2003}) that would allow removal of this source of $B$ modes with multi-frequency data.

\subsection{\texorpdfstring{$B$}{Lg}-mode signal due to patchy screening}
\label{subsec:screening}

Free electrons during the epoch of reionization act as a semi-opaque
screen between us and the CMB last-scattering surface.  The primary
anisotropies in the CMB are screened by an overall
factor of $e^{-\tau(\uvec{n},\chi_\ast)}$ as
\begin{align}
T(\uvec{n})&= e^{-\tau(\uvec{n},\chi_\ast)}T_{{\rm p}}(\uvec{n}), \\
(Q\pm iU)(\uvec{n})&= e^{-\tau(\uvec{n},\chi_\ast)}(Q\pm iU)_{\rm p}(\uvec{n}), 
\end{align}
where $T_{\rm p}$ and $(Q\pm iU)_{\rm p}$ are the primary (denoted by the
superscript $\rm{p}$) fluctuations in temperature and polarization,
respectively, which would be observed in the absence of reionization.
Variations in the optical depth due to the patchiness of reionization convert some of the primary $E$-mode into $B$-mode polarization. Linearising in the optical depth fluctuations, the power spectrum of these $B$ modes in the flat-sky limit is~\cite{DvorkinBmode}
\begin{align}
C_\ell^{BB\rm{(scr})}=e^{-2\bar{\tau}}\int
\frac{d^2\boldsymbol{\ell}_1}{(2\pi)^2}C_{\ell_1}^{EE\rm{(p)}}C^{\tau\tau}_{|\boldsymbol{\ell}-\boldsymbol{\ell}_1|}\sin^2 2(\phi_{\boldsymbol{\ell}_1}-\phi_{\boldsymbol{\ell}}) ,
\label{eq:clbbscr}
\end{align}
where
$C_{\ell_1}^{EE\rm{(p)}}$ is the primary $E$-mode power
spectrum and $\phi_{\ell_1}$ is the angle between $\boldsymbol{\ell}_1$ and the $x$-direction (which corresponds to $Q>0$).

The $E$-mode power peaks on small scales (around $\ell = 1000$), so large-scale $B$ modes from patchy screening are mostly sourced by modulation of these small-scale $E$-modes by small-scale fluctuations in the optical depth. The resulting screened signal behaves like white noise on large scales, with a constant power spectrum given by taking the limit $\ell_1 \gg \ell$ in the integrand of Eq.~\eqref{eq:clbbscr}~\cite{DvorkinBmode}:
\begin{equation}
C_\ell^{BB\rm{(scr})}\approx  e^{-2\bar{\tau}}\frac{1}{2}\int
\frac{d^2\boldsymbol{\ell}_1}{(2\pi)^2}C_{\ell_1}^{EE\rm{(p)}}C^{\tau\tau}_{\ell_1} .
\end{equation}
Despite employing the flat-sky approximation, Eq.~\eqref{eq:clbbscr} remains accurate even on large scales, where it recovers the correct white-noise amplitude as the full-sky expression in~\cite{DvorkinBmode}.

The total $B$-mode power spectrum induced by patchy reionization is
given by the sum of the quadrupole scattering and screening spectra:
\begin{equation}
C_\ell^{BB\rm{(tot})}=C_\ell^{BB\rm{( sca})}+C_\ell^{BB\rm{(scr})}.   
\end{equation}

\section{Reionization model calibrated to Lyman-\texorpdfstring{$\alpha$}{Lg} forest data}
\label{sec:modelling}

As a realistic model for the inhomogeneous distribution of free
electrons during the epoch of reionization, we consider the
post-processed cosmological radiation hydrodynamical simulation
presented by Kulkarni et al.\ \cite{2019MNRAS.485L..24K} as our
fiducial reionization model.  The underlying cosmological
hydrodynamical simulation is very similar in set-up to several
simulations from the Sherwood Simulation Suite
\cite{2017MNRAS.464..897B}.  This simulation was performed using the
\textsc{p-gadget-3} code, which is derived from the \textsc{gadget-2}
code \citep{2001NewA....6...79S, 2005MNRAS.364.1105S}.  The box size
used is $L=160 h^{-1}\,\text{cMpc}$.  The dark matter and gas distributions are
represented by $2048^3$ particles each.  The resultant dark matter
particle mass is $M_\mathrm{dm}=3.44\times 10^7 h^{-1}\,\text{M}_\odot$ while the
gas particle mass is $M_\mathrm{gas}=6.38\times 10^6 h^{-1}\,\text{M}_\odot$.
Periodic boundary conditions are imposed and the initial conditions
are set at $z=99$.  These are identical to the initial conditions of
the 160--2048 simulation from the Sherwood simulation suite
\cite{2017MNRAS.464..897B}.  These initial conditions are evolved down
to $z=4$. Snapshots of the gas density and other quantities are saved
at 40\,Myr intervals, which results in 38 snapshots.  We simplify galaxy
formation in this simulation by using the {\tt QUICK\_LYALPHA} option
in \mbox{\textsc{p-gadget-3}} to speed up the simulation. This removes
gas particles with temperature less than $10^5$\,K and overdensities greater than $1000$
from the hydrodynamical calculation by converting
them to star particles \citep{2004MNRAS.354..684V}. This approximation
does not affect the reionization process as the mean free path of
ionizing photons is determined by self-shielded regions with a typical
overdensity of $\Delta=10$--$100$ \citep{2009MNRAS.394.1812P,
  2018MNRAS.478.1065C}.  Before post-processing the results of the
hydrodynamic simulation, heat is injected in the gas distribution in
the simulation box assuming instantaneous reionization at redshift
$z=15$ and ionization equilibrium with the metagalactic UV
background modelled according to Haardt and Madau
\cite{2012ApJ...746..125H}, marginally modified to result in
inter-galactic medium (IGM) temperatures that agree with measurements
\cite{2011MNRAS.410.1096B}.  Note that we do not use the ionization
and temperature values that result out of this procedure; these values
are instead provided by the radiative transfer calculation that is
performed subsequently. The on-the-fly assumption of instantaneous
reionization is made simply to ensure that the gas distribution at
lower redshift has a realistic pressure smoothing, which plays a role
while calibrating the simulation to \lya\ data. An ideal approach
would be to perform the radiative transfer itself on the fly, but this
is still prohibitively expensive for the large dynamical ranges that
we seek here. Fortunately, the absence of coupling between the
radiative transfer and the hydrodynamic response to reionization
heating is unlikely to affect the calibration of the simulation to the
\lya\ forest data \cite{2018arXiv181011683O} as the pressure smoothing
scale at redshifts $z>5$ for our chosen UV background is less than $100h^{-1}\,\text{ckpc}$ \cite{2015ApJ...812...30K, 2017ApJ...837..106O},
approximately equal to the cell size of our grid (described below).
The hydrogen ionization solver assumes radiative cooling via two-body
processes such as collisional excitation of H~\textsc{i},
He~\textsc{i}, and He~\textsc{ii}, collisional ionization of
H~\textsc{i}, He~\textsc{i}, and He~\textsc{ii}, recombination, and
Bremsstrahlung \citep{1996ApJS..105...19K}, and inverse Compton
cooling off the CMB \citep{1986ApJ...301..522I}. Metal enrichment and
its effect on cooling rates is ignored.

We post-process the simulation with the radiative transfer code \textsc{ATON} 
\cite{2008MNRAS.387..295A, 2010ApJ...724..244A}.
In this approach, sources of ionizing radiation are assumed to be
present at the location of dark matter haloes in the simulation.  We
identify dark matter haloes in the output snapshots using the
friends-of-friends algorithm.  At $z=7$, the minimum halo mass in our
simulation is $2.3\times 10^{8}h^{-1}\,\text{M}_\odot$, which is close to
the atomic hydrogen cooling limit. The maximum halo mass at this
redshift is $3.1\times 10^{12}h^{-1}\,\text{M}_\odot$. We assume that a
halo with mass $M$ emits hydrogen-ionizing photons at a rate $\dot
N=\alpha M$, where $\alpha$ is a free parameter that encodes our
ignorance of the complex astrophysical processes, such as star
formation and interaction with the inter-stellar medium, which govern
the production of ionizing photons from galaxies. Note that the
average ionizing photon emissivity of the simulated volume is then
$\dot n = \alpha \sum M / V_\mathrm{box}$ where $V_\mathrm{box}=L^3$ is
the volume of the simulation box and the summation is over all haloes.
The parameter $\alpha$ is a function of redshift but is independent of
halo mass. It is the only parameter that is varied in order to
calibrate the simulation to given observations, such as the
\lya\ forest \cite{2019MNRAS.485L..24K}.  While performing the
radiative transfer, we place sources only in haloes with masses
greater than $10^9\,$M$_\odot$ as the simulated halo mass function
below this mass suffers from incompleteness due to lack of resolution.
Further, as \textsc{aton} solves the radiative transfer equation on a
Cartesian grid, we project the smooth particle hydrodynamic (SPH)
kernels of the gas particles in our simulation onto such a grid.  We
choose the number of grid cells equal to the number of gas particles
in the simulation, which yields a grid resolution of
$78.125h^{-1}\,\text{ckpc}$. The \textsc{aton} code uses a moment-based
description with the M1 approximation \cite{2008MNRAS.387..295A} for
the Eddington tensor to solve the radiative transfer equation. In
order to reduce computational cost, we use a single photon frequency
for the radiative transfer.  We assume that all sources have a
blackbody spectrum with $T=70\,000\,\text{K}$ \cite{2018MNRAS.477.5501K}. This
corresponds to an average photon energy of 23.83\,eV in the optically-thick limit. The reionization history and the calibration with the
\lya\ forest are both robust to variations in the assumptions
regarding the source spectrum and photon frequency. A change in the
source spectrum has the effect of changing the resultant temperature
of the gas distribution in the simulation.  However, any resultant
deviations in the \lya\ forest can be compensated by changing $\alpha$
in the source emissivity above.

\begin{figure*}
  \begin{center}
    \begin{tabular}{cc}
      \includegraphics[width=0.48\textwidth]{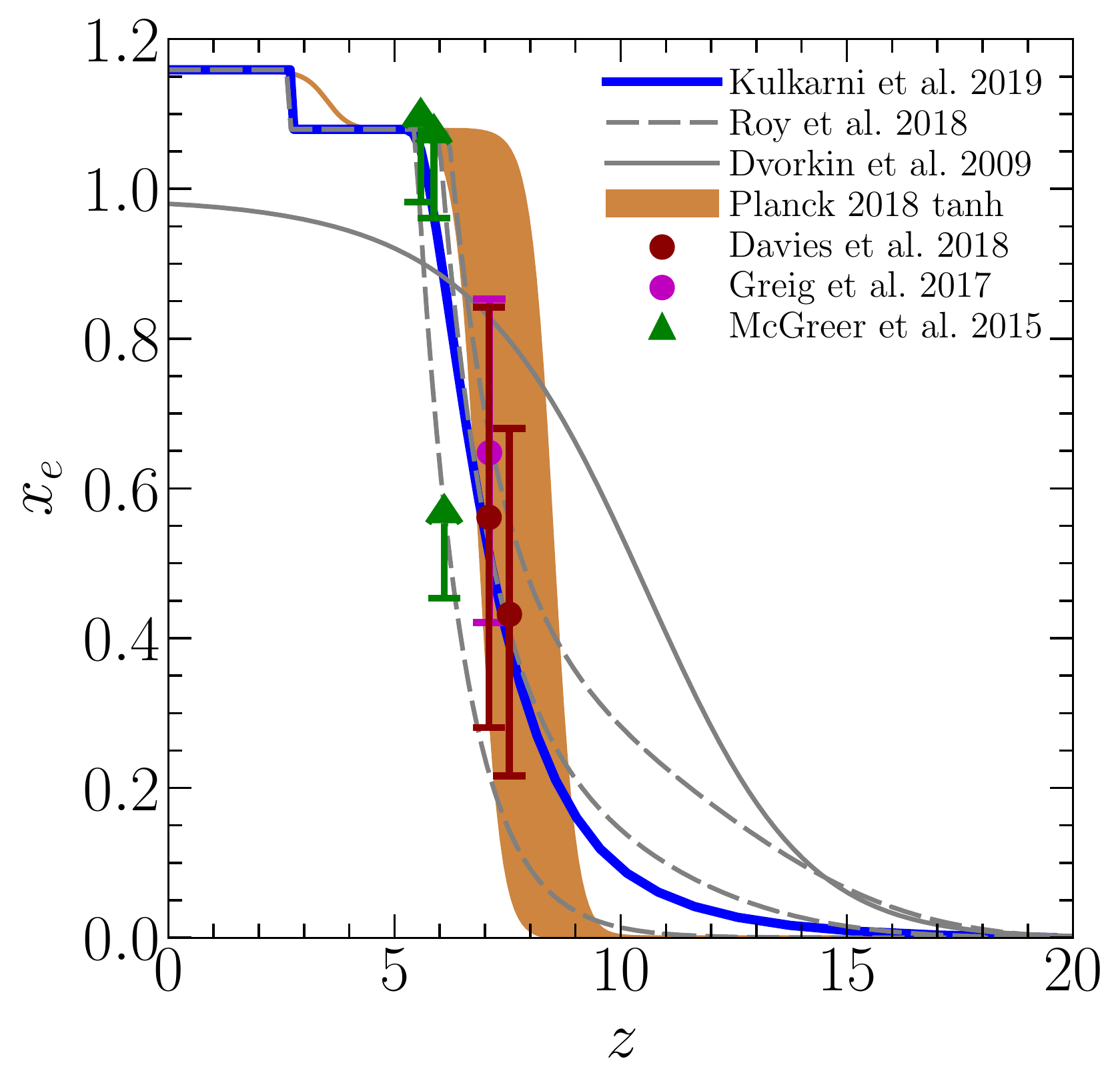} & 
      \includegraphics[width=0.48\textwidth]{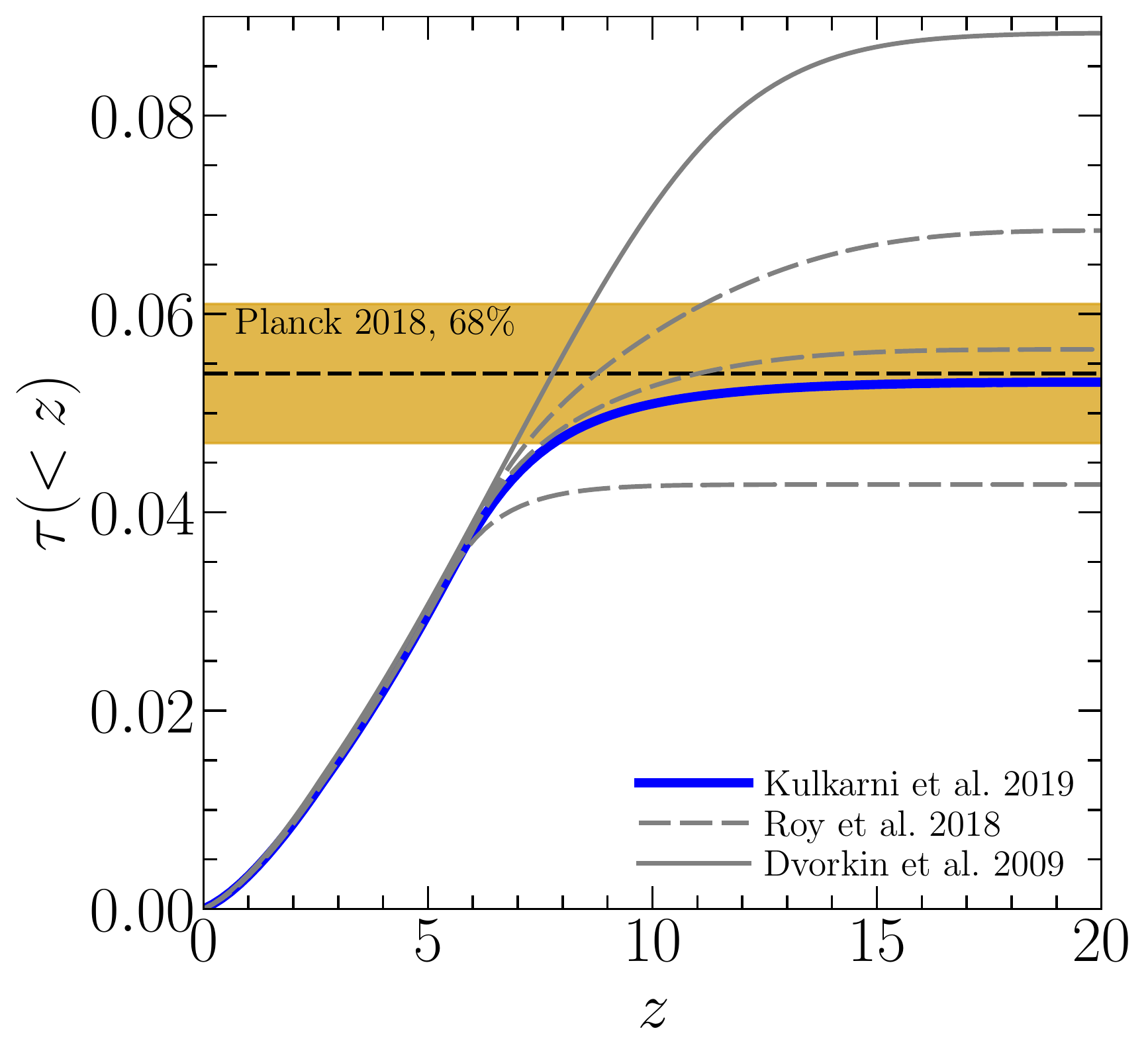}
    \end{tabular}
    \caption{\emph{Left}: Evolution of the free-electron fraction in
      the radiative transfer simulations of Kulkarni et
      al.\ \cite{2019MNRAS.485L..24K} that we consider as our fiducial
      reionization model in this work (blue lines).  Also shown are
      constraints from the spectra of two highest-redshift quasars
      \cite{2017MNRAS.466.4239G, 2018ApJ...864..142D} and from the
      incidence of dark pixels in Ly$\alpha$ and Ly$\beta$ forests
      \cite{2015MNRAS.447..499M}.  Grey curves show reionization
      histories considered in earlier work \cite{Roy2018, Lapi2017,
        Dvorkin2009} for comparison.  The brown shaded region shows
      the 68\,\% confidence region of the tanh reionization model
      reported by Planck~\cite{Planck2018}.  \emph{Right}: Cumulative
      electron scattering optical depth in the reionization models
      considered in the left panel, compared with the constraint from
      Planck of $\tau = 0.054\pm 0.007$ (68\,\%
      confidence)~\cite{Planck2018}.
      \label{fig:xe_comparison}}
  \end{center}
\end{figure*}

The calibration of our simulation with \lya\ data is demonstrated and
discussed in Ref. \cite{2019MNRAS.485L..24K}. Here we discuss the
reionization history implied by this model. The left panel of
Figure~\ref{fig:xe_comparison} shows the evolution of the
volume-averaged free electron fraction $x_\mathrm{e} =
n_e/n_p$ in our simulation compared with other
models and data. In our model, hydrogen reionization completes rather
late, at $z=5.2$. The universe is half ionized at $z_\mathrm{re}=7$.
We can quantify the duration of reionization $\Delta z$ as the
difference between the redshifts at which the universe is 5\,\% and 95\,\%
reionized. In our model, we have $\Delta z=5.4$.  As
Figure~\ref{fig:xe_comparison} shows, this reionization history is
more gradual than the tanh models considered in the Planck 2018
results \cite{Planck2018}. While reionization in our simulations
begins earlier compared to the best-fit tanh models from Planck 2018,
reionization also ends somewhat later. This reionization history is
consistent with recent constraints from the spectra of two
highest-redshift quasars \cite{2017MNRAS.466.4239G,
  2018ApJ...864..142D} and the relatively model-independent
constraints from the incidence of dark pixels in Ly$\alpha$ and
Ly$\beta$ forests \cite{2015MNRAS.447..499M}. Note that as shown in
Figure~\ref{fig:xe_comparison}, we assume an instantaneous
reionization of He~\textsc{ii} at $z=2.7$ \cite{2016ApJ...825..144W,
  2018arXiv180805247W}. We have taken into account the electrons generated during the first reionization of helium, which adds a further 8\,\% of electrons beyond those from hydrogen reionization.
The right panel of
Figure~\ref{fig:xe_comparison} shows the average cumulative Thomson
scattering optical depth out to $z=20$ in our simulation. It is found
to agree very well with the measurement reported by Planck 2018
\cite{Planck2018} of $\tau = 0.054\pm 0.007$ (68\,\% confidence).

In Figure~\ref{fig:xe_comparison}, we also show the reionization histories adopted in previous work by some of the current authors~\cite{Roy2018}. These are derived (see Section 2 of their paper for details)
by solving the standard ionization/recombination
balance equation for the spatially-averaged ionization fraction, and
exploits the cosmic star-formation rate density by \cite{Lapi2017} based on the most
recent determination of the dust-corrected UV, far-IR, and radio
luminosity functions out to high redshift. In Figure
\ref{fig:xe_comparison} (left panel) we plot the reionization
histories corresponding to a conservative value $f_{\rm esc}=10\,\%$ of
the escape fraction of ionizing photons from primeval galaxies, and to
three different values $M_{\rm UV}=-17$, $-13$, and $-12$ for the faintest
UV magnitudes contributing to the ionizing background.
The reionization history and cumulative optical depth for $M_{\rm UV}\approx -13$ is close to that of our fiducial radiative transfer simulation.
The key difference between these models and our simulation results is
that the spatial distribution of the
free electrons should be more realistic in the latter (which follows radiative transfer).
This is important for the derived ionization power spectra and the induced $B$-mode power spectra, which we discuss below.

The box size $L$ determines the largest scale accessible in our
reionization model. This corresponds to the fundamental mode of the
box, which is given by $k_\mathrm{f}=2\pi/L$. For the fiducial
simulation described above the box size is $L=160 h^{-1}\,\text{cMpc}$, which implies
$k_\mathrm{f}=0.039h\, \text{cMpc}^{-1}$. The value of $k_\mathrm{f}$ determines
the minimum multipole, $\ell_{\rm min}$, we can calculate the angular power
spectra at as we have $\ell \approx k_{\mathrm{f}}\chi(z)$. This allows us to reach down to
$\ell_{\rm min}\approx 430$ in the $L=160 h^{-1}\, \text{cMpc}$ radiative transfer
simulations. However, the angular power spectrum of the primordial $B$-mode signal
peaks around $\ell \sim 100$.  To assess the contribution from patchy reionization on such scales, we
need to extend the size of the simulation. It is computationally prohibitive to increase the box size $L$ significantly in the radiative transfer simulations. Instead, we adopt a well-known semi-numerical method
based on the excursion set approach to modelling reionization.  We
validate this method by matching its reionization history to that of our radiative transfer simulation and 
by comparing the ionization power spectra on scales where there is overlap.
In this way, we use the
semi-numerical method to model the free electron distribution 
during the epoch of reionization up to scales of $1\,\text{cGpc}$,
which corresponds to a minimum multipole $\ell_\mathrm{min}\approx 50$, below the multipole where the primordial signal is expected to peak. The simulation with $L=1\,\text{cGpc}$ has the same halo mass resolution as the radiative transfer simulation of reionization. 

We use the 21cmFAST code \cite{Mesinger2011, Mesinger2007} to generate
the density field from redshift $z_{\rm start}=19.5$ to $z_{\rm
  end}=4.97$ at $40\,\text{Myr}$ intervals. Gaussian initial conditions are
set on a $3584^3$ grid at $z\approx 300$. 21cmFast then calculates the linear
displacement field on a coarser $512^3$ grid using the Zel'dovich
approximation \cite{Zeldovich1970, 1985ApJS...57..241E}.  These
displacements are then used to update the particle positions in the
initial density field.  This method of generating large-scale density
fields using the Zel'dovich approximation is computationally efficient
and has been validated by several works \cite[e.g.,][]{Davies2017,
  Davies2016, Santos2009, Mesinger2007}. The coarse resolution of the
grid is sufficient for our purposes as we are primarily interested in
the large-scale distribution of ionized hydrogen. Next, we identify
haloes with masses above $10^9\,\text{M}_\odot$ using the extended
Press--Schechter formalism \cite{1993MNRAS.262..627L,
  1991ApJ...379..440B}.  21cmFAST finds haloes iteratively
by smoothing the density field at progressively smaller scales and
identifying haloes of a given mass as regions where the smoothed overdensity in the (Zel'dovich-evolved) field first exceeds that
for gravitational collapse, $\delta_c=1.686$.
A real space top-hat filter is used for smoothing the
density field and assigning masses to the haloes.

\begin{figure*}
  \begin{center}
    \includegraphics[width=0.6\textwidth]{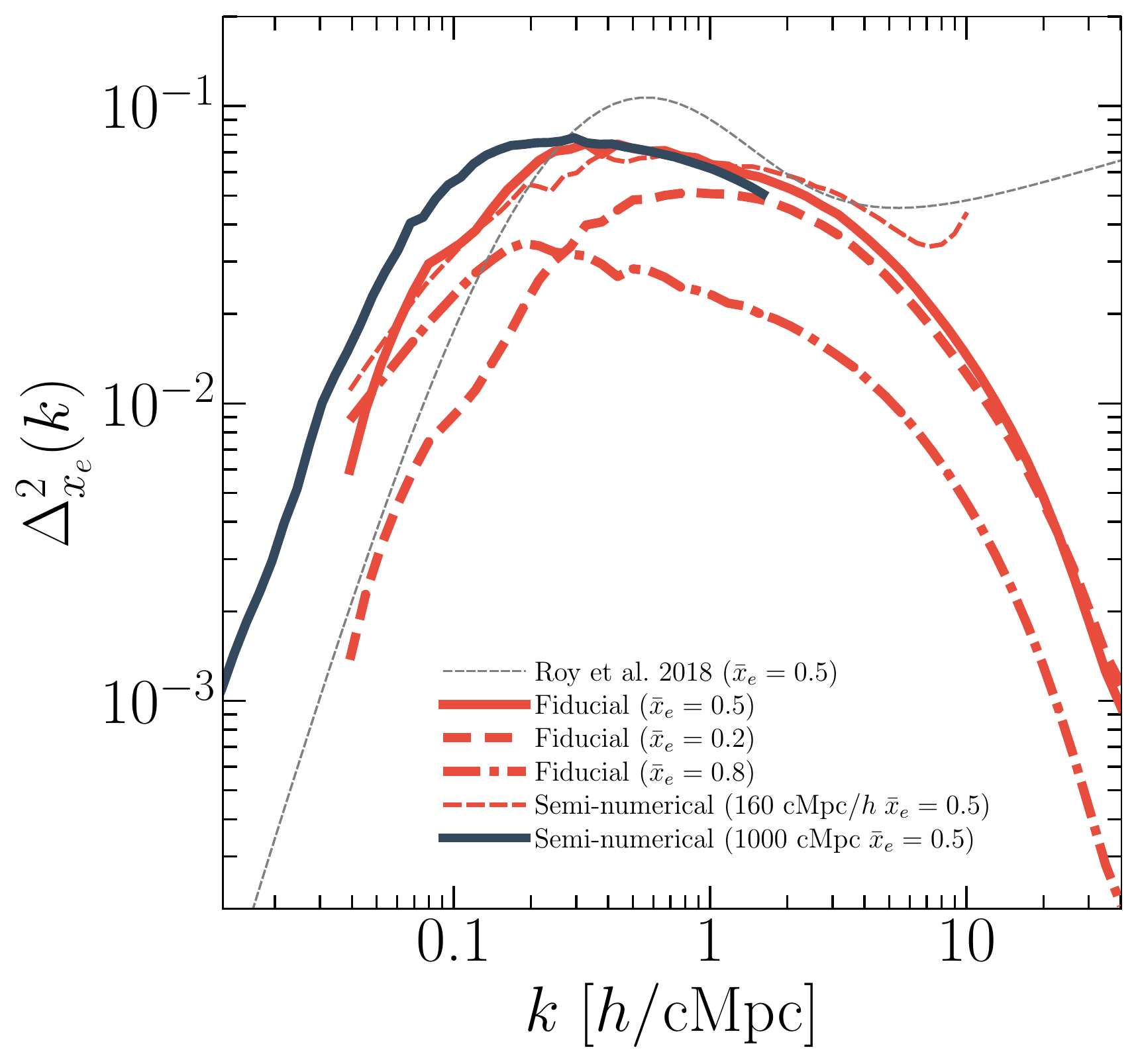}
    \caption{Dimensionless power spectra of the ionized hydrogen fraction from our fiducial radiative transfer simulation (thick red curves) at redshifts corresponding to average ionization fractions of $\bar{x}_e = 0.8$ (dot-dashed), $0.5$ (solid) and $0.2$ (dashed).
    The thin-dashed red curve shows
      the same for $\bar{x}_e=0.5$ from the semi-numerical method
      implemented in a $L=160 h^{-1}\, \text{cMpc}$ long box and the thick black curve for $L=1\,\text{cGpc}$.
      The power spectrum at $\bar{x}_e=0.5$ from a semi-analytic method used in recent work by some of the authors~\cite{Roy2018} is shown for comparison (thin grey dashed).}
    \label{fig:pxe_sherwood}
  \end{center}
\end{figure*}

Given the density field and haloes, we derive the ionization field by
placing sources of Lyman-continuum radiation in dark matter haloes and
using the excursion-set method \citep{2004ApJ...613...16F,
  2009MNRAS.394..960C, 2011MNRAS.411..955M}.  Similar to our radiative
transfer simulation above, the total number of ionizing photons
$N_{\gamma}$ produced by a halo is assumed to be proportional to the
halo mass \citep{2016MNRAS.463.2583K}:

\begin{equation}
  N_\gamma(M) = N_\gamma^\mathrm{LyC}M / m_p,
  \label{eqn:ngammam}
\end{equation}
where the dimensionless proportionality factor $N_\gamma^\mathrm{LyC}$ includes the
Lyman-continuum escape fraction and $m_p$ is the proton mass.
A grid cell at position $\vec{x}$
is ionized if the condition
\begin{equation}
  \langle n_\gamma(\vec{x})\rangle_R > \langle
  n_p(\vec{x})\rangle_R(1+\bar N_\mathrm{rec}),
  \label{eqn:exset1}
\end{equation}
is satisfied in a spherical region centred on the cell for some radius
$R$ \citep{2004ApJ...613....1F, 2009MNRAS.394..960C,
  2011MNRAS.411..955M}. Here, the averages are over the spherical
region, $n_p$ is the number density of hydrogen nuclei,
\begin{equation}
  n_\gamma = \int_{M_\mathrm{min}}^\infty
  dM\left.\frac{dN}{dM}\right\vert_{R}N_\gamma(M),
  \label{eqn:ngamma}
\end{equation}
where $dN/dM|_R$ is the halo mass function within the spherical
region, $M_\mathrm{min}$ is the minimum halo mass that contributes
ionizing photons, and $\bar N_\mathrm{rec}$ is the average number of
recombinations per hydrogen atom in the IGM. (We note that the global
volume average of $n_p$ is $\bar{n}_{p}$.)  The condition in
Eq.~\eqref{eqn:exset1} simply requires that a sufficient number of photons have been produced in a spherical region of radius $R$ to have ionized all the hydrogen atoms in that region, and to have maintained this ionization state in the presence of $\bar{N}_\text{rec}$ recombinations per atom. It can be recast as
\begin{equation}
  \zeta_\mathrm{eff}f(\mathbf{x},R)\geq 1,
  \label{eqn:exset}
\end{equation}
where the quantity
\begin{equation}
  f=\rho_m(R)^{-1}\int_{M_\mathrm{min}}^\infty
  dM\left.  \frac{dN}{dM}\right\vert_{R}M, \label{eqn:fcoll}
\end{equation}
is the collapsed fraction into haloes of mass $M>M_\mathrm{min}$ within the spherical region centred on $\vec{x}$,
$M_\mathrm{min}$ is the minimum mass of halos that emit Lyman
continuum photons, and $\rho_m(R)$ is the average matter density within the sphere.
The parameter $\zeta_\mathrm{eff}$ quantifies the number of photons in the
IGM per hydrogen atom in stars, accounting for hydrogen recombinations
in the IGM. We can write $\zeta_\mathrm{eff}$ in terms of the
parameters of Eqs~\eqref{eqn:ngammam} and \eqref{eqn:exset1} as
\begin{equation}
  \zeta_\mathrm{eff} =
  \frac{N^\mathrm{LyC}_\gamma}{1-Y_\mathrm{He}}(1+\bar
  N_\mathrm{rec})^{-1},
\end{equation}
where $Y_\mathrm{He}$ is the helium mass fraction. The parameter
$\zeta_{\text{eff}}$ is the only parameter that determines the
ionization field in this approach. The volume-weighted ionized
fraction in the simulation box is $Q_V\equiv
\sum_iQ_i/n_\mathrm{cell}$, where the ionized volume fraction in a
cell $i$ is $Q_i$ and $n_\mathrm{cell}$ is the total number of grid
cells. We tune $\zeta_\mathrm{eff}$ to ensure that our
$1\,\text{cGpc}$ box has the same reionization history as the
radiative transfer simulation described above.

We derive the 3D power spectrum of the electron fraction by averaging $|\Delta x_e(\vec{k})|^2$, where
$\Delta x_e (\vec{k})$ is the Fourier transform of the free electron fraction, within a spherical shell of radius $k$.
Figure~\ref{fig:pxe_sherwood} shows the dimensionless power spectrum
$\Delta_{x_e}^2(k)=k^3P_{x_e x_e}(k)/(2\pi^2)$ at three different redshifts in
our fiducial radiative transfer simulation. The power spectrum peaks when the
global ionization fraction is close to 50\,\%, before this time there are few free electrons and later the IGM is fully ionized so the electron fraction is smooth.
In the same figure we also compare
the power spectrum from our fiducial simulation with an implementation of
the semi-numerical scheme in a box of the same size ($L=160 h^{-1}\,\text{cMpc}$ on a side). This
semi-numerical implementation is tuned to have an identical
reionization history as our fiducial radiative transfer simulation by
choosing an appropriate value for the $\zeta_\mathrm{eff}$ parameter
as described above. We find that the power spectrum of the electron
fraction from the semi-numerical approach agrees very well with that from
the radiative transfer simulation. The difference in power is less than 10\,\% on
most scales. This validates the semi-numerical approach and our
calibration of it. To extend beyond the lowest wavenumber $k$ that we can probe, we proceed to consider the semi-numerical scheme in larger boxes with $L=500 h^{-1}\,\text{cMpc}$ (not shown in Figure~\ref{fig:pxe_sherwood}) and $L=1000\,\text{cMpc}$. These larger simulations
are also tuned to have the same reionization history as the fiducial radiative transfer simulation.
They have very similar power as the fiducial simulation on small scales
but additional power on large scales (see Figure~\ref{fig:pxe_sherwood})\footnote{As we use a fixed number of grid points, we cannot measure the power spectrum in the large box to such small scales as in the fiducial simulations.}.
This additional power arises from
ionized regions around rare, high-mass haloes that are not present in the
smaller box.
The agreement on small scales is good because
we use the same minimum halo mass value of $10^9\,\text{M}_\odot$ in every
semi-numerical simulation. For comparison with earlier work, we include in Figure~\ref{fig:pxe_sherwood} the power spectrum at $\bar{x}_e = 0.5$ from a semi-analytical model of reionization used in Ref.~\cite{Roy2018}. 
One-bubble correlations
dominate on scales $k\geq 1\,\text{cMpc}^{-1}$ while large scales are dominated by the
correlation between different ionized bubbles. Note that the semi-numerical scheme is known to display a lack of numerical convergence in the ionization power spectrum due to non-conservation of photons \cite{Choudhury:2018kae, 2007ApJ...654...12Z}.  It is important to note, therefore, that the choice of our semi-numerical model is based on its agreement with the radiative transfer simulation at the overlapping scales and therefore with the data.

\section{Results}
\label{sec:results}
\subsection{Power spectrum of secondary \texorpdfstring{$B$}{Lg}-mode anisotropies}

\begin{figure*}
  \begin{center}
    \begin{tabular}{cc}
      \includegraphics[width=0.48\textwidth]{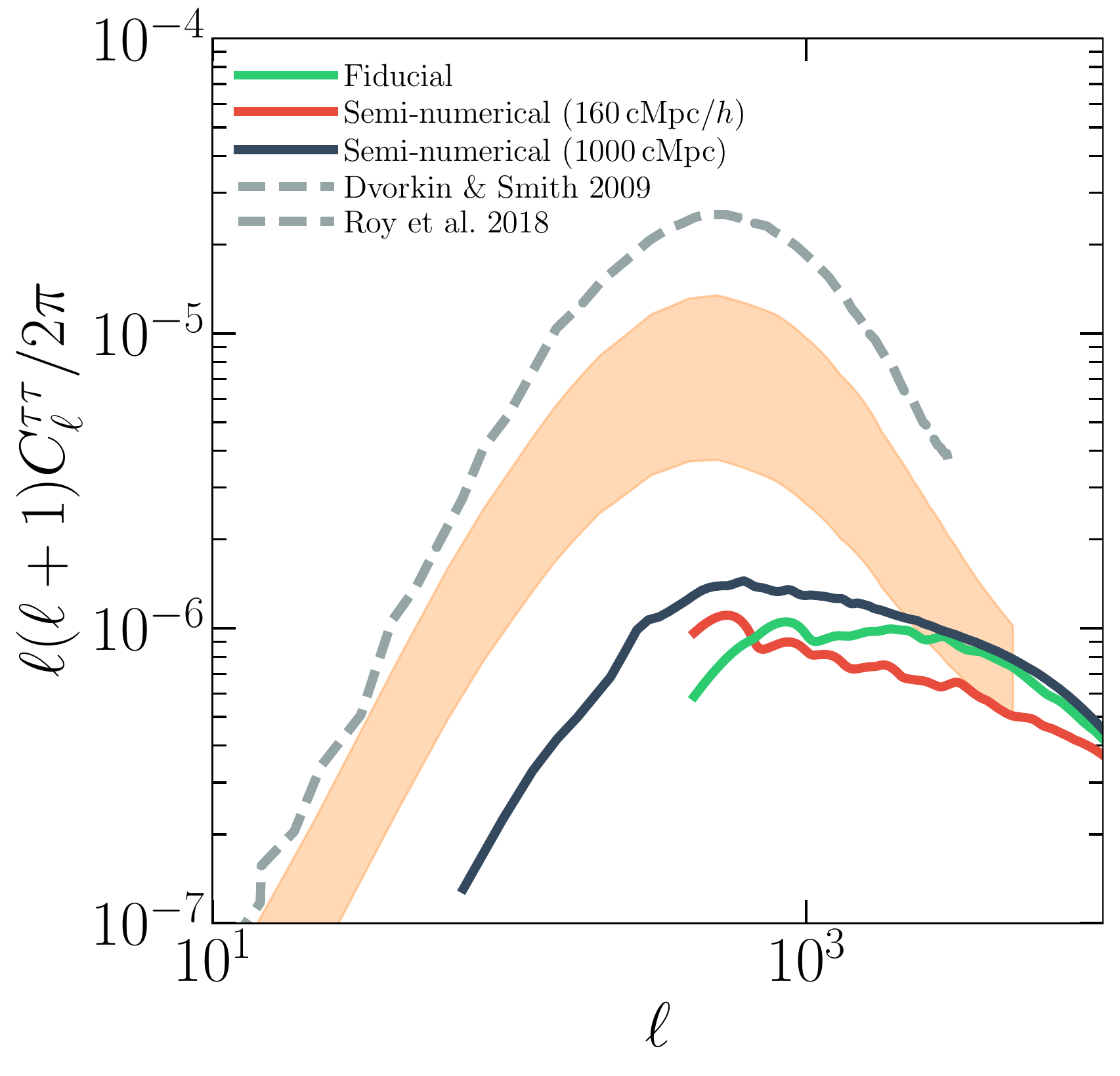} &
      \includegraphics[width=0.48\textwidth]{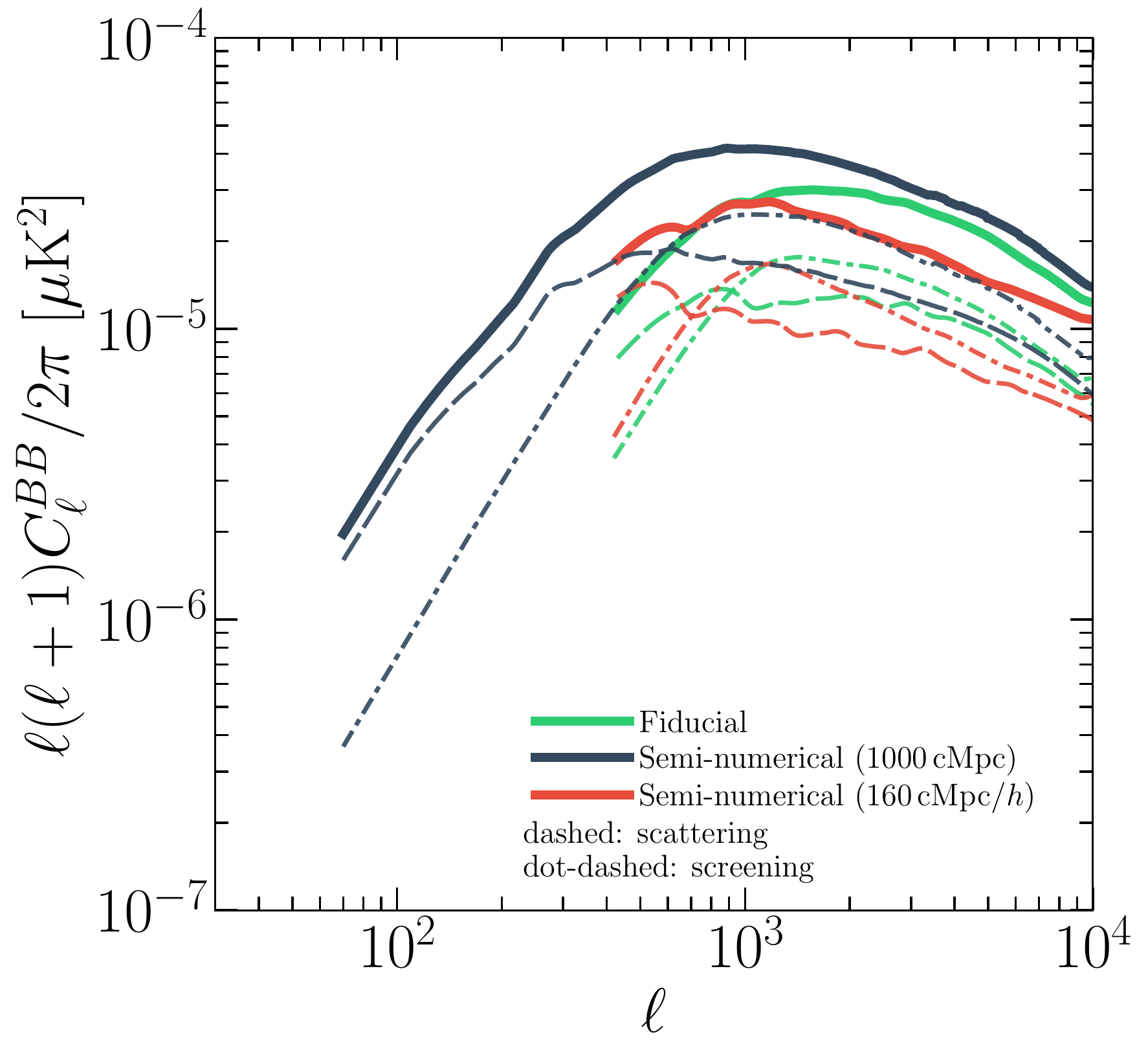}
    \end{tabular}
    \caption{\emph{Left}: angular power spectrum of the Thomson
      scattering optical depth in our fiducial radiative transfer
      simulation (green curve), semi-numerical simulation in the $160
      h^{-1}\,\text{cMpc}$ box (red curve), and the semi-numerical
      simulation in the $1\,\text{cGpc}$ box (black curve).  All
      three simulations have the same reionization histories.  The
      dashed grey curve in this panel shows the power spectrum from
      \cite{Dvorkin2009} for comparison. The shaded region corresponds
      to the range of power spectra derived previously \cite{Roy2018}.
      \emph{Right}: $B$-mode power spectrum from patchy reionization,
      with the same color coding as the left panel. The dashed and
      dot-dashed curves show the quadrupole scattering and screening
      contributions, respectively, to the power spectrum and the solid
      lines are their total.
      \label{fig:cltau_and_clbb}}
  \end{center}
\end{figure*}

In this section we estimate the $B$-mode signal from patchy
reionization and calculate its contamination in the search for the
primordial signal. Figure \ref{fig:cltau_and_clbb} (left panel)
compares the angular power spectrum of the Thomson optical depth
fluctuations from the fiducial radiative transfer simulation to those
from the semi-numerical simulations using the excursion set algorithm
(calibrated to the radiative transfer simulation, as described in
Section~\ref{sec:modelling}).  The optical depth power spectrum
$C_\ell^{\tau \tau}$ agrees reasonably well at multipoles $\ell
\gtrsim 700$ for the various simulations. (Some level of disagreement
is expected as the radiative transfer is much better at computing the
ionization field around the edges of the bubble
\cite{2007ApJ...654...12Z}.)  The figure also compares our simulation
results to earlier work by other authors~\cite{Dvorkin2009}. The power
spectrum $C_\ell^{\tau\tau}$ from the $1\,\text{cGpc}$ box is
roughly $20$ and $15$ times smaller than the fiducial model studied by
\cite{Dvorkin2009} at $\ell\approx 100$ and $\ell\approx 1000$,
respectively. The same toy model of reionization is used by
Refs.~\cite{Dvorkin2009} and \cite{Roy2018} but their reionization
histories are different. The optical depth used in \cite{Dvorkin2009}
is higher than \cite{Roy2018}, hence $C_\ell^{\tau\tau}$ is larger.

The right-hand panel of Figure~\ref{fig:cltau_and_clbb} shows the $B$-mode angular power spectrum
due to patchy reionization. The $B$ modes from quadrupole scattering (dashed lines) exceed 
those from patchy screening (dot-dashed lines) on large scales, $\ell\lesssim 600$, across all our
simulations. As noted in Section~\ref{subsec:screening}, the 
$B$ modes from patchy screening are sourced by a modulation of the
primary $E$ modes and have white-noise power on scales large compared to the
peak of the $E$-mode power spectrum ($\ell \approx 1000$). The large-scale $B$-mode power is determined by the
small-scale fluctuations in the optical depth, where the $E$-mode power is significant.
In contrast, large-angle $B$ modes
due to quadrupole scattering depend on the large-scale power of
$C_\ell^{\tau\tau}$. Since the latter is dominated by the shot noise of the bubbles, the $B$-mode power from
scattering is also white on scales large compared to the typical bubble size. 

The total $C_\ell^{BB}$ across our different simulations peaks in the range $\ell = 1000$--$2000$ (in the remainder of this paper, $C_\ell^{BB}$ will always refer to this total $B$-mode signal 
from patchy reionization, i.e., the sum of scattering and screening contributions).
The shape of the angular power spectrum due to scattering closely follows $C_\ell^{\tau\tau}$ (see Eq.~\ref{eq:BBscatter}). It peaks on larger scales than the screening contribution, as the latter turns over near the peak of the primary $E$-mode spectrum. Scattering contributes 80\,\% of the total $B$-mode power from patchy reionization at $\ell=100$, with the remaining 20\,\% from patchy screening. Note that the authors of~\cite{2019arXiv190301994M} do not include the screening contribution in their estimates. Although the total patchy $B$-mode power calculated here agrees with~\cite{2019arXiv190301994M} to around 5\,\%, their scattering contribution is 24\,\% higher than our estimates. The maximum amplitude of the total $C_\ell^{BB}$ derived from the semi-numerical method for the $L=1\,\text{cGpc}$ box is around twice that of our fiducial radiative-transfer simulation.
 This is due to the presence of large bubbles around massive haloes that are present in the
$L=1\,\text{cGpc}$ box but are absent in the smaller $L=160h^{-1}\,\text{cMpc}$ box.

\begin{figure}
  \begin{center} 
    \includegraphics[width=0.9\textwidth]{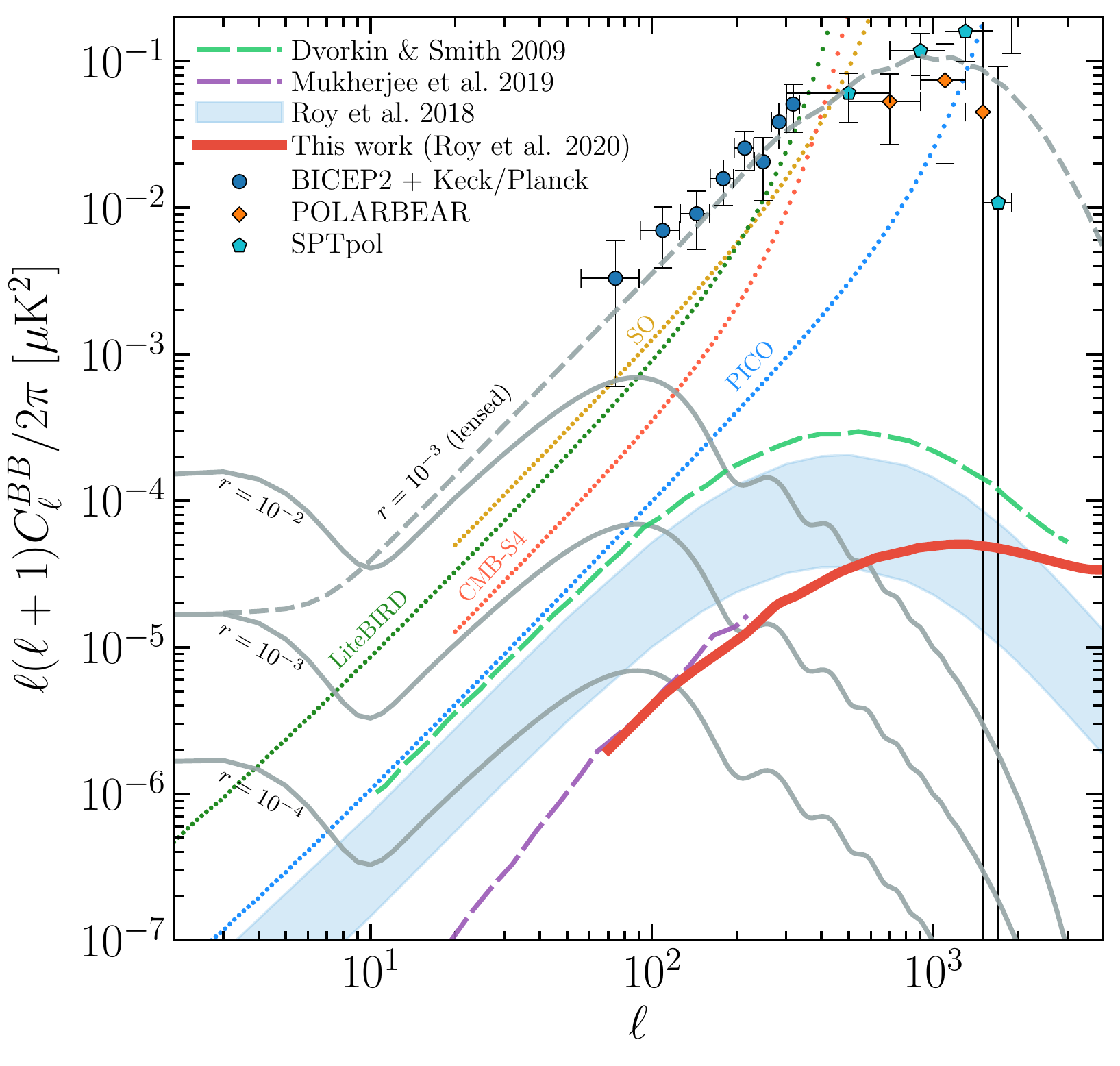}
  \end{center}
  \caption{Comparison of the $B$-mode power spectrum from patchy
    reionization (red solid curve) for our $L= 1\,\text{cGpc}$
    semi-numerical simulation with the primordial $B$-mode spectrum
    (solid grey curves) for tensor-to-scalar ratios $r=10^{-2}$,
    $10^{-3}$ and $10^{-4}$. Also shown are previous estimates of the
    reionization-induced signal \cite{Roy2018, Dvorkin2009,
      2019arXiv190301994M}. Dotted curves show the instrumental noise
    power spectra for PICO (blue), CMB-S4 (red), LiteBIRD (green), and
    Simons Observatory (yellow). The sum of the primordial and
    lensing-induced $B$-mode power is shown for $r=10^{-3}$ as the
    dashed grey curve. Data points show $B$-mode power measurements
    from BICEP2/Keck Array~\cite{2015PhRvL.114j1301B} (after
    foreground cleaning), POLARBEAR~\cite{2017ApJ...848..121P} and
    SPT~\cite{2015ApJ...807..151K}.}
\label{fig:clbball}
\end{figure}

In Figure~\ref{fig:clbball} we compare the $B$-mode angular power spectrum from patchy reionization to the
primordial $B$-mode power from gravitational waves and the instrumental noise power spectra expected
for forthcoming and proposed experiments. The noise spectra are modelled as
$N_\ell^{BB}=\Delta_P^2\exp[\ell(\ell+1)\Theta_f^2/(8\ln 2)]$
\cite{1995PhRvD..52.4307K}. Here, $\Delta_P$ is the polarization noise level in the maps and
$\Theta_f$ is the full-width half-maximum (FWHM) of the beam.
We consider four experimental configurations; two ground-based and two satellite missions. For the ground-based experiments we consider
$\Delta_P = 3.0\,\mu\text{K-arcmin}$ and $\Theta_f=17\,\text{arcmin}$, corresponding to the goal white-noise level at $145\,\text{GHz}$ for the Simons Observatory small-aperture telescope (SAT) survey~\cite{SOforecast}, and $\Delta_P = 1.5\,\mu\text{K-arcmin}$ and $\Theta_f=30\,\text{arcmin}$, corresponding roughly to the planned SAT survey from CMB-S4~\cite{2019arXiv190704473A, CMBS42016}.
The combination of limited sky coverage and low-frequency atmospheric noise will limit the largest scales accessible to these ground-based surveys ($\ell \sim 30$ in polarization).
For the full-sky satellite surveys, we consider 
$\Delta_P = 2.4\,\mu\text{K-arcmin}$ and $\Theta_f=30\,\text{arcmin}$, corresponding to the total sensitivity of the LiteBIRD satellite~\cite{Hazumi2019}.
More sensitive satellite missions have also been proposed; here we consider PICO~\cite{PICO2019} with 
$\Delta_P = 0.85\,\mu\text{K-arcmin}$ and $\Theta_f=7.9\,\text{arcmin}$.
Several ground-based instruments, such as AdvACT~\cite{AdvACT2016}, SPT-3G~\cite{2014SPIE.9153E..1PB}, 
BICEP3~\cite{BICEP32014}, and Simons Array~\cite{Simonsarray2016}, are
operating at present but with lower sensitivity than the noise curves shown in 
Figure~\ref{fig:clbball}. Future
space CMB missions will aim to measure the primordial $B$-mode signals
from both (global) reionization at 
$\ell\leq20$ and recombination, which peaks around $\ell \approx 100$,
while ground-based surveys will mostly target the recombination signal.
In Figure~\ref{fig:clbball}, we show three primordial $B$-mode power spectra
corresponding to tensor-to-scalar ratios $r=10^{-2}$, $10^{-3}$ and $10^{-4}$. The current best constraint on
$r$ comes from joint analysis
of $B$-mode data from BICEP/Keck Array, Planck and WMAP: $r< 0.06$
at 95\,\% confidence \cite{2018PhRvL.121v1301B}.

CMB photons are also lensed due to the gravitational potential of
matter along the line of sight. Primary $E$ modes are transformed into
lensed $B$ modes, even if there is no primordial $B$-mode signal. On the large scales of interest, the $B$ modes due to lensing can be considered as white noise with a
corresponding $\Delta_P\approx 5\,\mu\text{K-arcmin}$.
Future (ground-based) CMB experiments will need to remove
the lensing $B$-mode signal using \enquote{delensing} techniques to
probe primordial gravitational waves for $r\leq 10^{-3}$. The sum of the primordial and lensing $B$-mode power is also shown in Figure~\ref{fig:clbball} for $r=10^{-3}$.

The $B$-mode power from patchy reionization shown in Figure~\ref{fig:clbball} (red line) is from
the semi-numerical method in the larger $L=1\,\text{cGpc}$ box. As discussed in Section~\ref{sec:modelling}, this should be our most reliable estimate of the large-scale power.
We compare this with the power estimated in previous studies using a semi-analytic model~\cite{Hu2006,Dvorkin2009,Roy2018}.
These semi-analytic studies predict a larger $B$-mode power than our simulations.
We note that the reionization history used in Ref.~\cite{Dvorkin2009} was based on WMAP data for
which $\tau=0.089$, partly accounting for the increased $B$-mode power (see Section~\ref{sec:history}).
A further relevant factor is the 
larger characteristic bubble radius used in the earlier works ($R_b\approx 5\,\text{cMpc}$).
Typically, the $B$-mode signal due to
patchy reionization peaks at a scale comparable to $R_b$,
with larger characteristic radius shifting the peak in the power spectrum to smaller angular multipoles.
In our simulation-driven results, the peak power is around $\ell \sim 1000$ (close to the peak in the lensing power), whereas for the bubble radii adopted in earlier semi-analytic models the peak is in the range $\ell = 100$--$500$.
Our estimate of the large-scale $B$-mode power is consistent with that in recent work~\cite{2019arXiv190301994M} using
photon-conserving semi-numerical simulations (dashed purple line in Fig.~\ref{fig:clbball}) and a reionization history consistent with the 
central value of the optical depth from Planck~\cite{Planck2018} ($\tau\approx 0.055$).
We compare with their results for a minimum halo mass $M_{\mathrm{min}}=10^9$ $M_{\odot}$, consistent with the choice in our simulations.
While Ref.~\cite{2019arXiv190301994M} considered only the
quadrupole scattering contribution, we also include screening. However, on large scales,
$\ell \lesssim 100$, the $B$ modes from scattering dominate.

It is clear from Fig.~\ref{fig:clbball} that patchy reionization will have a limited impact on searches for primordial $B$-modes from the next generation of surveys that are targeting $r \gtrsim 10^{-3}$. However, it may become relevant for $r \lesssim 10^{-4}$. (We discuss these issues more quantitatively in Section~\ref{sec:fisher}.)
Given the shape of the signal from patchy reionization, it is a negligible contaminant in searches for $B$ modes from gravitational waves sourced by homogeneous scattering at reionization (i.e., at $\ell \lesssim 20$), targeted by future space missions such as the LiteBIRD and PICO~\cite{Hazumi2019,PICO2019}.

\subsection{Effect of reionization history}
\label{sec:history}
\begin{figure*}
  \begin{center}
    \begin{tabular}{cc}
      \includegraphics[width=0.48\textwidth]{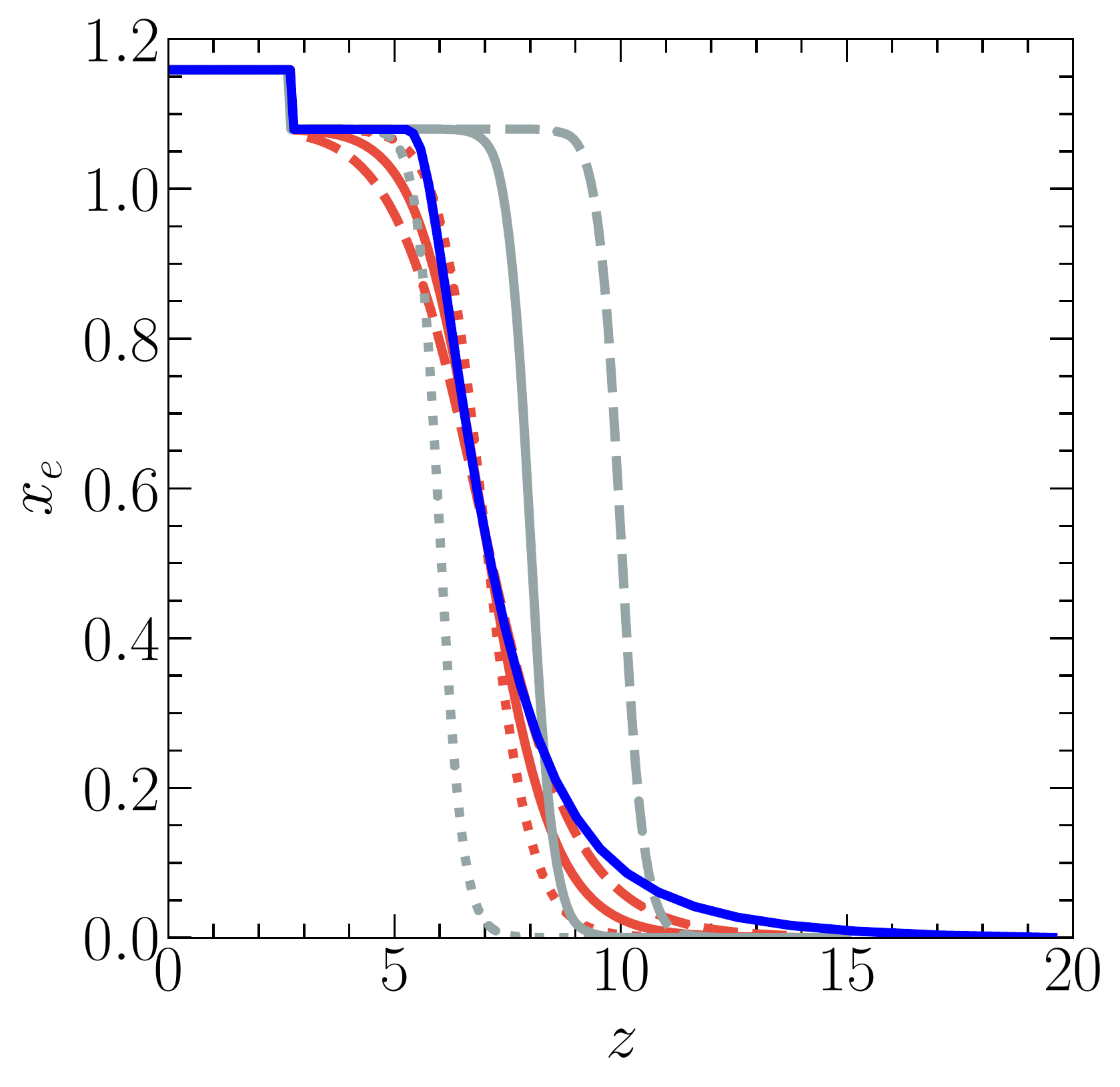} & 
      \includegraphics[width=0.48\textwidth]{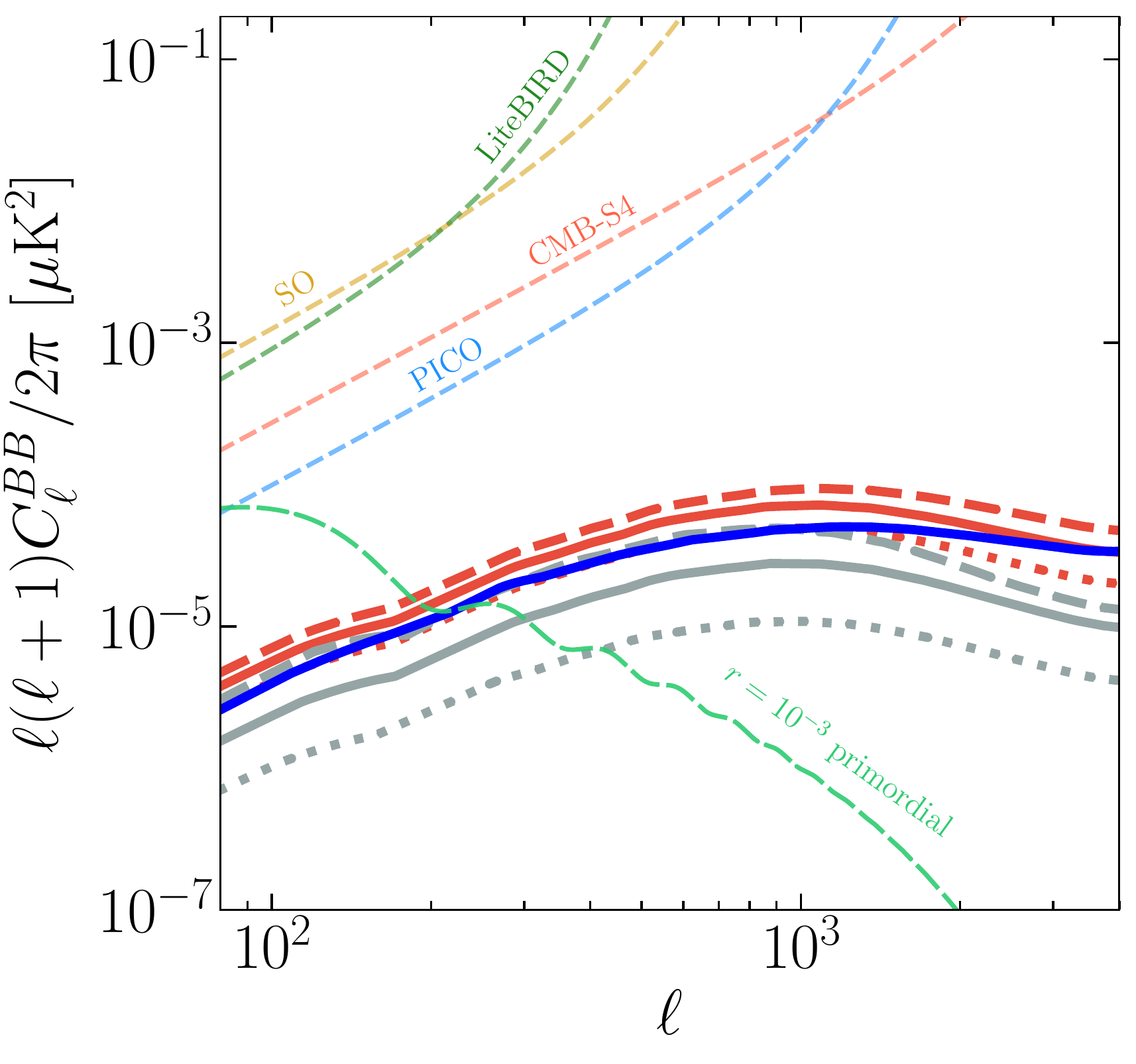}
    \end{tabular}
    \caption{\emph{Left}: Reionization histories for the tanh
      parameterization of Eq.~\eqref{eqn:tanh} compared to our
      fiducial model (blue). The red curves show the effect of varying
      the reionization duration $\Delta z_{\rm re}=1$ (dotted), $1.5$ (solid) and $2$ (dashed)
      at fixed mean redshift $z_{\rm re} = 7$. The grey curves instead
      vary $z_{\rm re} = 6$ (dotted), $8$ (solid) and $10$ (dashed) for fixed $\Delta z_{\rm re}
      = 0.5$.  \emph{Right}: Corresponding $B$-mode power spectra
      calculated with our semi-numerical method, with the same colour
      coding as the left panel. The amplitude of the $B$-mode power is
      higher for larger $z_{\rm re}$ and $\Delta z_{\rm re}$.  The
      dashed curves show noise power spectra for the same surveys as
      in Figure~\ref{fig:clbball} and the primordial signal for the
      tensor-to-scalar ratio $r=10^{-3}$ (green).
\label{fig:clbb_tanh}}
  \end{center}
\end{figure*}

Within our semi-numerical framework we can easily vary the reionization history to explore its impact on the $B$-mode power from patchy reionization. We adopt a simple
$\tanh$ functional behaviour, parameterized by
the mean redshift of reionization, $z_{\rm re}$, and
the width of the reionization, $\Delta z$.  The ionization fraction
evolves as
\begin{equation}
  x_e(z)=\frac{f}{2}\left[1+\tanh\left(\frac{y_{\rm re}-y}{\Delta y_{\rm re}}\right)\right],
  \label{eqn:tanh}
\end{equation}
where $y(z)=(1+z)^{3/2}$, $y_{\rm re}=y(z_{\rm re})$ and $\Delta
y_{\rm re}=1.5\sqrt{1+z_{\rm re}}\Delta z_{\rm re}$~\cite{Lewis:2008wr}. 
Accounting for the first reionization of helium, we have $f \approx 1.08$.
In the left panel of Figure~\ref{fig:clbb_tanh} we show example reionization histories varying 
$z_{\rm re}$ and $\Delta z_{\rm re}$. We note that some of these histories would be excluded by constraints on $\tau$ from Planck, but are included to illustrate more obviously the impact of the reionization history on the $B$-mode power from patchy reionization.

In the right panel of Figure~\ref{fig:clbb_tanh} we show the $B$-mode power from patchy reionization computed within our semi-numerical framework for the reionization histories shown in the left panel. The amplitude of the $B$-mode power spectrum increases for earlier reionization and with the duration $\Delta z_{\rm re}$.  For the duration, the main effect is just that more variance is accumulated because of the increased duration of the patchy phase. For the timing, the signal increases due to the higher density if reionization happens earlier.

\subsection{Minimum mass of reionization sources}
\label{subsec:minmass}

In our fiducial simulation, sources of reionizing radiation are placed
in haloes with masses down to $10^9\,\text{M}_\odot$.  The mass of
reionization sources will have an effect on the clustering of the free
electron distribution during reionization \cite{2019MNRAS.485.3486D,
  2017MNRAS.469.4283K}, which can in turn change the induced secondary
$B$ modes.  The agreement of our radiative transfer simulation with a
variety of data \cite{2019MNRAS.485L..24K, 2019arXiv190512640K}
suggests that sources in haloes down to $10^9$~M$_\odot$ should be
able to reionize the universe.  Still, the mass of reionizing sources
is unfortunately unknown.  The fraction of hydrogen-ionizing radiation
produced in a galaxy that can escape into the intergalactic medium and
cause reionization is largely uncertain.  This escape fraction has
been measured in a handful of relatively bright and low-redshift
galaxies \cite{2010ApJ...725.1011V, 2018A&A...616A..30C,
  2018arXiv180601741F, 2018ApJ...869..123S, 2017MNRAS.468..389J,
  2016ApJ...826L..24S, 2015ApJ...810..107M, 2014Sci...346..216B,
  2016A&A...585A..48G, 2016Natur.529..178I, 2018arXiv180511621M,
  2018MNRAS.478.4851I, 2018MNRAS.474.4514I, 2018A&A...616A..30C} but
is unknown for the galaxies within the epoch of reionization.  The
escape fraction can be a function of halo mass and can influence, e.g., the
relative contribution of highly-clustered massive haloes and low-mass
haloes to reionization.  Furthermore, active galactic nuclei (AGN), which
are known to have unit escape fraction, can also reionize the universe
instead of star-forming galaxies.  Such a scenario, although unlikely
\cite{2018arXiv180709774K}, can potentially have a large effect on the
induced $B$-mode signal as AGN reside in very high mass haloes that
are highly clustered \cite{2017MNRAS.469.4283K}.

\begin{figure}
\centering
 \includegraphics[width=0.6\textwidth]{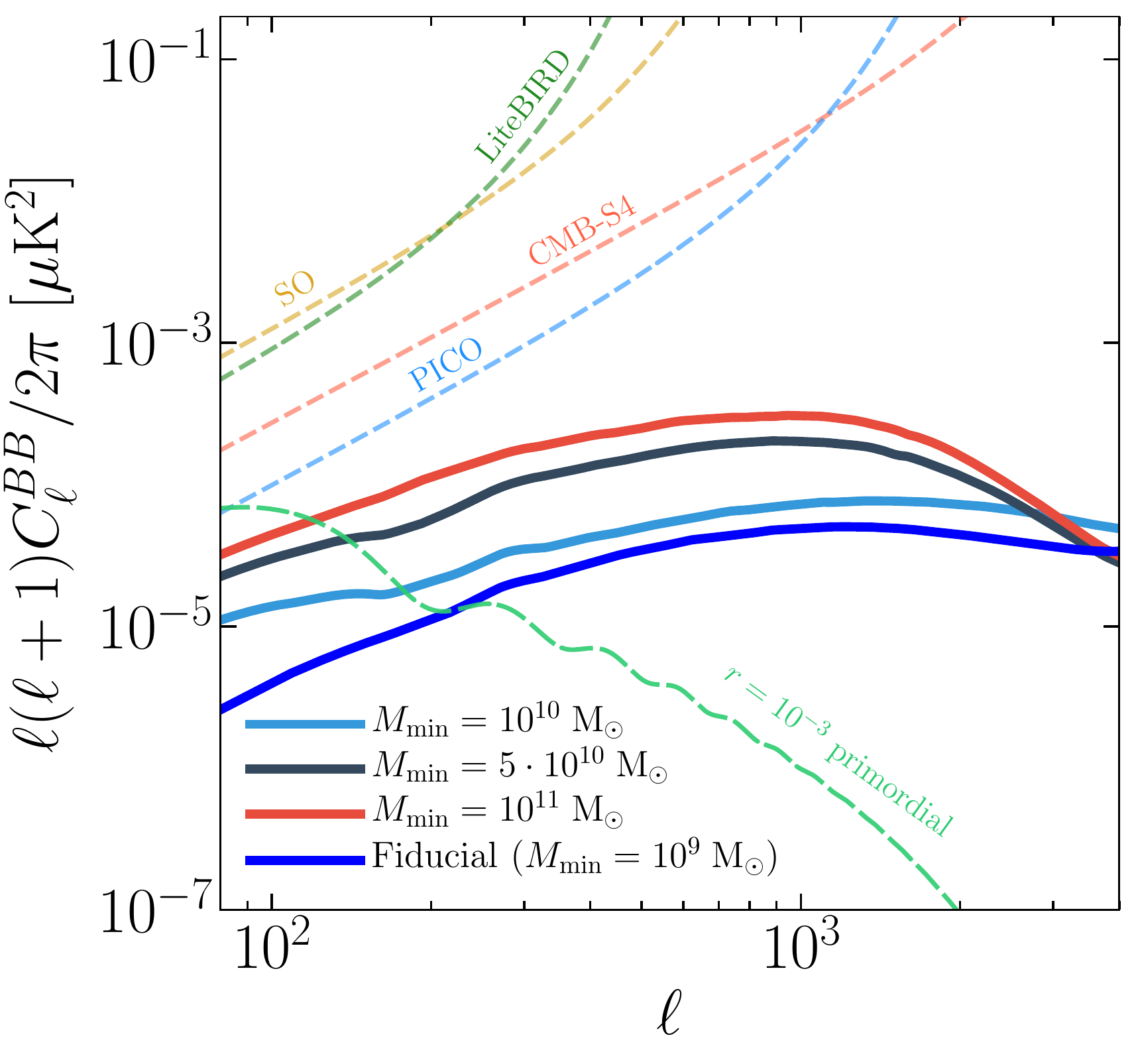}
\caption{Dependence of the induced $B$-mode power spectrum on the
  minimum mass of reionizing sources for a fixed reionization history.
  Sources in higher-mass haloes induce larger $B$-mode power due to
  their stronger clustering.  The dashed curves show noise power
  spectra for the same surveys as in Figure~\ref{fig:clbball} and the
  primordial signal for the tensor-to-scalar ratio $r=10^{-3}$
  (green).  }
\label{fig:clbb_mmin}
\end{figure}

In Figure~\ref{fig:clbb_mmin}, we increase the minimum mass of
reionizing sources in our model from $M_{\rm min} = 10^9\,\text{M}_\odot$ to $10^{11}\,\text{M}_\odot$.  This is done using the excursion-set-based
treatment described in Section~\ref{sec:modelling}.
The reionization history is always kept
the same as our fiducial simulation (Figure~\ref{fig:xe_comparison})
by changing the brightness of each halo suitably, i.e., by tuning the
parameter $N_\gamma^\mathrm{LyC}$ in Eq.~(\ref{eqn:ngammam}).  As a
result, as $M_{\rm min}$ is increased the ionized regions increase in
size and become more clustered \cite{2017MNRAS.469.4283K}.  This
results in an enhancement of about an order of magnitude in the
induced $B$-mode power spectrum, as seen in
Figure~\ref{fig:clbb_mmin}.

\subsection{Constraints on the tensor-to-scalar ratio}\label{sec:fisher}

We now quantitatively consider the effect of reionization-induced
secondary $B$ modes on the inference of the tensor-to-scalar ratio
from the $B$-mode power spectrum.  

The primordial $B$-mode power spectrum peaks around $\ell \approx
100$, where the signal is generated by scattering at recombination.
Current and future CMB polarization experiments aim to target this
recombination peak in the search for primordial gravitational
waves. For small values of the tensor-to-scalar ratio $r$, the
signal-to-noise ratio (S/N) is expected to be less than one per mode
(see Figure~\ref{fig:clbball}), but such small signals can still be
detected if the information across multiple modes is
combined~\cite{SOforecast, CMBS42016, 2018SPIE10698E..4FS,
  2018SPIE10698E..1YS}.  Also, foreground contamination due to
emission from the Galaxy exceeds the primordial signal at the
recombination peak in even the cleanest regions of the sky for
$r=10^{-2}$ \cite{1995PhRvD..52.4307K, 2016JCAP...03..052E,
  2016A&A...588A..65K, 2015PhRvL.114j1301B}.  As a result, the precise
shape of the $B$-mode power spectrum will be hard to determine for
small $r$ and it will be a challenge to distinguish the primordial
signal from secondary $B$ modes induced by other sources if the shape of
the secondary $B$-mode power spectrum is not sufficiently dissimilar
within the range of multipoles considered.

Let us begin by assuming that no attempt is made to account for the
additional $B$-mode power from patchy reionization when constraining
$r$. Generally, we can calculate the bias introduced by
unaccounted-for secondary anisotropies as
\begin{equation}
\Delta r = \left(\sum_{\ell = \ell_{\text{min}}}^{\ell_{\text{max}}} \left[C_\ell^{BB}(r=1)\right]^2 / \Delta C_\ell^2\right)^{-1} 
\sum_{\ell = \ell_{\text{min}}}^{\ell_{\text{max}}} C_\ell^{BB\text{(sec)}} C_\ell^{BB}(r=1) / \Delta C_\ell^2 .
\end{equation}
Here, $C_\ell^{{BB}\text{(sec)}}$ is the secondary $B$-mode power from
non-primordial sources such as lensing, patchy reionization, and
foreground emission, and $C_\ell^{BB}(r=1)$ is the primary power for
$r=1$. The minimum and maximum multipoles used in the analysis are
denoted by $\ell_{\text{min}}$ and $\ell_{\text{max}}$,
respectively. We focus on multipoles targeted by ground-based
experiments around the peak of the primary power from recombination
and assume $\ell_{\text{min}}=70$ and $\ell_{\text{max}}=200$. The
variance of the total $B$-mode power is $\Delta C_\ell^2$ which is approximately $2C_\ell^2/[(2\ell+1)f_{\text{sky}}]$, where $f_{\text{sky}}$ is the fraction of sky in the survey. 

\begin{figure}
\centering
\includegraphics[width=0.8\textwidth]{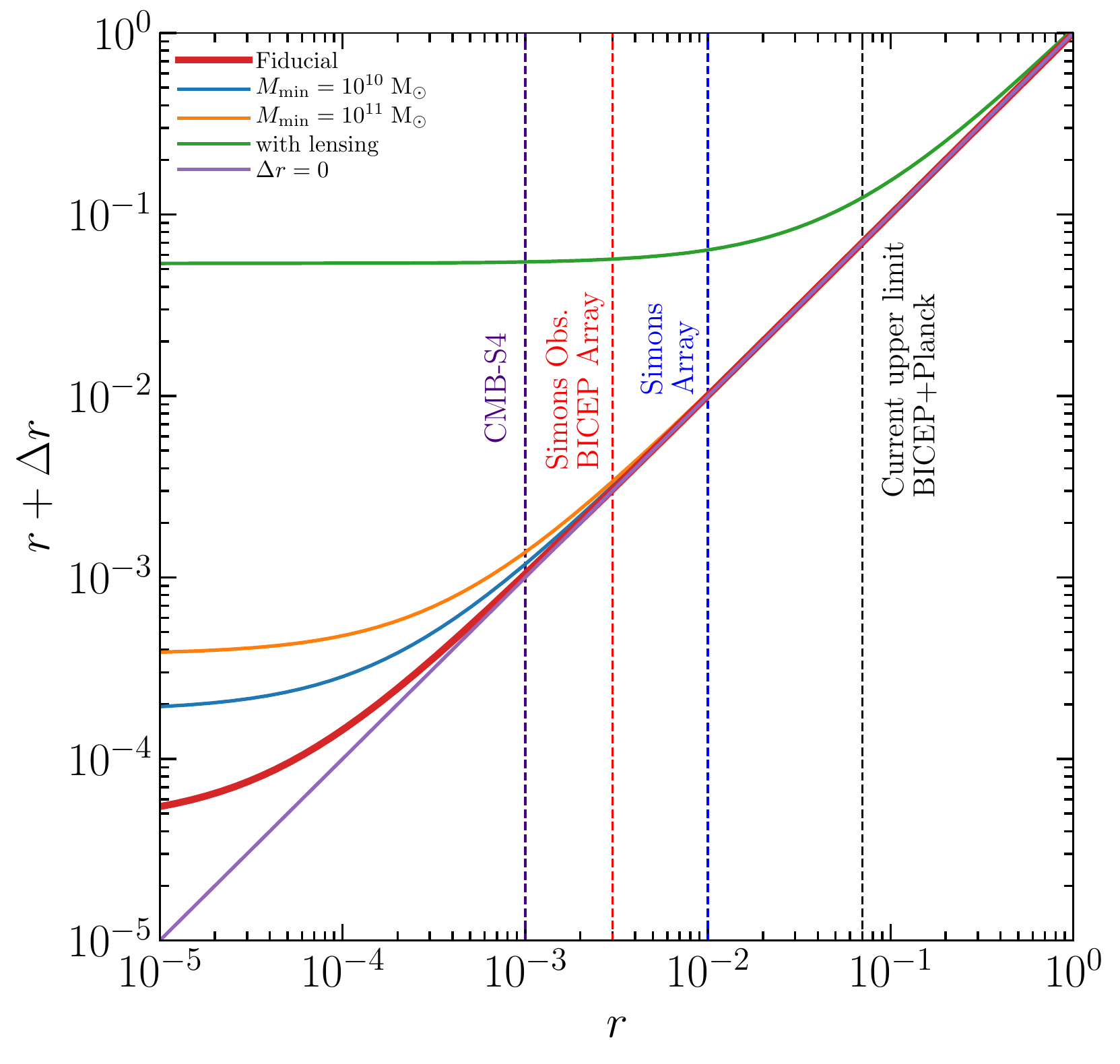}
\caption{Bias on the tensor-to-scalar ratio inferred from the $B$-mode
  power spectrum in the multipole range $70\leq \ell \leq 200$ when
  secondary sources of $B$-mode power are not accounted for.  The red
  curve corresponds to the case where the reionization-induced
  $B$-mode signal is the only unaccounted-for secondary signal.  Blue and
  orange curves show how this bias increases with the minimum mass of halos
  hosting ionizing sources.  The green curve shows the bias if gravitational
  lensing is also not accounted for. Vertical lines show the current upper limit
  on $r$ and predicted experimental sensitivities for Simons Array,
  BICEP Array~\cite{2018SPIE10708E..07H}, Simons
  Observatory~\cite{2019JCAP...02..056A}, and
  CMB-S4~\cite{2016arXiv161002743A}.}
\label{fig:reff}
\end{figure}

Figure~\ref{fig:reff} shows the bias on $r$ for our fiducial
reionization model, and the effect of increasing the minimum halo mass
hosting ionizing sources.
Secondary $B$ modes from patchy reionization can bias the
tensor-to-scalar ratio $r$ for $r < 10^{-3}$. At the sensitivity level
of future ground-based experiments ($r$ around $10^{-3}$),
reionization-induced $B$ modes could generate a fractional bias
between $10\,\%$ and $30\,\%$ for
$M_{\text{min}}=10^{9}\,\text{M}_\odot$ (our fiducial model) and
$10^{11}\,\text{M}_\odot$, respectively.  For a future satellite
experiment (e.g., PICO \cite{2018SPIE10698E..4FS} or CMB Bharat
\cite{CMB-Bharat-Proposal}) targeting $r=5\times 10^{-4}$, the bias
could be in the range $35$--$75\,\%$ as $M_{\rm min}$ varies from
$10^{9}\,\text{M}_\odot$ to $10^{11}\,\text{M}_\odot$, respectively,
if these experiments used only the recombination signal.  However,
these space-based missions will constrain $r$ by probing both the
recombination and reionization signal. The latter will be negligibly
biased by reionization for detectable levels of $r$.

We find that the fractional bias on $r$ due to the $B$ modes induced by patchy reionization agrees with the findings of \cite{2019arXiv190301994M}, as expected given that the amplitude of the induced $C_\ell^{BB}$ is of the same order as reported there (see Figure~\ref{fig:clbball}). Recall, however, that Ref.~\cite{2019arXiv190301994M} do not include the signal from patchy screening. In the earlier work of Mortonson \& Hu~\cite{Mortonson2006}, the amplitude of the $B$-mode power is around an order of magnitude higher than estimates here (see Table 1 there) for $\tau=0.06$, and would result in a similarly higher bias in $r$.
This is due to the size distribution of ionized bubbles being very different there compared to what we see in our simulations.

In practice, future CMB experiments will search for the primordial $B$-mode signal in the presence of the lensing signal, instrumental noise and foregrounds. The lensing-induced $B$ modes can be partly removed using high-resolution polarization measurements over an overlapping sky region along with some estimate for the lensing deflections, obtained either from the CMB itself or in combination with other correlated tracers, in a process known as delensing. The residual $B$-mode power after delensing, and any residual foregrounds after cleaning with multi-frequency data, will likely be dealt with by marginalsing over the amplitudes of suitable template power spectra as part of the likelihood analysis used to estimate the tensor-to-scalar ratio. In such an analysis, we would model the $B$-mode spectrum as
\begin{equation}
C_\ell^{BB} = \sum_{i} \alpha_i C_\ell^{BB(i)} \, ,
\label{eq:BBmodel}
\end{equation}
where $\alpha_i$ are the parameters we wish to constrain jointly, with
$\alpha_1 = r$ and the other parameters are amplitude parameters for
different sources of secondary power (e.g., residual lensing and patchy
reionization). The template spectrum for the $i$th parameter is
$C_\ell^{BB(i)}$, which reduces to $C_\ell^{BB}(r=1)$ for $i=1$ and to
the appropriate secondary power template for $i>1$. 
The maximum-likelihood estimates for the parameters are
\begin{equation}
\hat{\alpha}_i = \sum_j \left(F^{-1}\right)_{ij} \sum_{\ell = \ell_{\text{min}}}^{\ell_{\text{max}}} C_\ell^{BB(j)} \hat{C}_\ell^{BB} / \Delta C_\ell^2 \, ,
\end{equation}
where $\hat{C}_\ell^{BB}$ is the measured power spectrum (with any instrument noise bias subtracted) and
the Fisher matrix for the parameters is
\begin{equation}
F_{ij} = \sum_{\ell = \ell_{\text{min}}}^{\ell_{\text{max}}}   C_\ell^{BB(i)} C_\ell^{BB(j)} / \Delta C_\ell^2 \, .
\end{equation}
After marginalising over the amplitudes of the secondary spectra, the error on the tensor-to-scalar ratio is $\left( F^{-1}\right)_{11}$.

In the forecasts below, we assume polarization noise $\Delta_P = 1.5\,\mu\text{K-arcmin}$ and a beam size $\Theta_f = 30\,\text{arcmin}$. We further assume that the lensing power has been reduced by 90\,\% through delensing \cite{2019arXiv190704473A}. We set the sky fraction to $f_{\text{sky}}=1$, noting that statistical errors on parameters scale as $f_{\text{sky}}^{-1/2}$ while biases are independent of $f_{\text{sky}}$. In this experimental configuration, we find that the statistical error on $r$ is $\sigma(r)=1.3\times 10^{-4}$ without marginalising over the amplitude of the patchy $B$-mode power from reionization. The bias on $r$ from patchy reionization ranges from $5.5\times 10^{-5}$ to $6.3\times 10^{-4}$ as the minimum halo mass ranges from $M_{\text{min}}=10^9\,\text{M}_\odot$ to $M_{\text{min}}=10^{11}\,\text{M}_\odot$, assuming no correction is made. If instead we marginalise over patchy reionization, with a template corresponding to our fiducial model with minimum halo mass $M_{\text{min}}=10^9 \text{M}_\odot$ (and no prior on the amplitude), we find that $\sigma(r)$ increases to $2\times 10^{-4}$. In Table \ref{tab:bias}, we show the increase in $\sigma(r)$ after marginalising over the amplitude of the patchy $B$-mode power from reionization for several future ground-based and space-based CMB experiments.

The maximum-likelihood parameter estimates are unbiased if the templates have the correct shape: $\langle \hat{\alpha}_i \rangle = \alpha_i^{\text{true}}$, where $\alpha_i^{\text{true}}$ are the true parameter values.
However, if there is some error in the $I$th template, so the true power in Eq.~(\ref{eq:BBmodel}) has an additional contribution $\delta C_\ell^{BB(I)}$, the parameter means become
\begin{equation}
\langle \hat{\alpha}_i \rangle = \alpha_i^{\text{true}} + \sum_j \left(F^{-1}\right)_{ij} \sum_{\ell = \ell_{\text{min}}}^{\ell_{\text{max}}} C_\ell^{BB(j)} \delta C_\ell^{BB(I)} / \Delta C_\ell^2 \, .
\end{equation}
We note that only the $I$th parameter mean changes if $\delta C_\ell^{BB(I)} \propto C_\ell^{BB(I)}$, corresponding to a change in amplitude of the $I$th source of power. However, if there is an error in the \emph{shape} of the template, this template mismatch can bias constraints on $r$. We assess this level of bias by considering an extreme case where the true model has a minimum halo mass $M_{\text{min}}=10^{11}\text{M}_\odot$, but the template assumed in the analysis has $M_{\text{min}}=10^9 \text{M}_\odot$. Without marginalisation, the bias on $r$ is $6.3\times 10^{-4}$, as discussed above. However, with marginalisation, although with an incorrect template, the bias falls significantly to $9.9\times 10^{-5}$, which is around half the (marginalised) full-sky statistical error on $r$.

We have not assumed any prior on the amplitude of the power from patchy reionization in these forecasts. Given our current theoretical uncertainties -- for example, over the relevant multipole range the maximum ratio of power in the models with $M_{\text{min}}=10^9 \text{M}_\odot$ and $M_{\text{min}}=10^{11} \text{M}_\odot$ is around 13 -- the $B$-mode measurements should be sufficiently informative to differentiate between our fiducial and more extreme models for the patchy power, and so adopting a prior that brackets the theoretical uncertainty would give little improvement in the marginalised statistical error on $r$. In fact, adopting such a prior risks exacerbating the bias in $r$ since both template mismatch and errors in the central value of the prior can lead to bias in this case.
For reference, in our fiducial $M_{\text{min}}=10^9 \text{M}_\odot$ case, the marginalised error on the amplitude of the template is $\sigma(\alpha)=2.7$, which is insufficient to detect the power in the fiducial model ($\alpha=1$) but would rule out the $M_{\text{min}}=10^{11} \text{M}_\odot$ model at high significance.

\begin{table}[tb]
\centering


\begin{tabular}{lccc}
\hline\hline

&  & \multicolumn{2}{c}{With marginalisation} \\
\cline{3-4}
Experiment & $\sigma(r)$ & $\sigma(r)$ & $\sigma(r)$ \\
& & $M_{\text{min}}=10^9 \text{M}_\odot$ & $M_{\text{min}}=10^{11} \text{M}_\odot$ \\
\hline
CMB-S4 & $1.3\times 10^{-4}$ &$2.0\times 10^{-4}$ & $2.4\times 10^{-4}$  \\ 
Simons Observatory &  $3.9\times 10^{-4}$ &$5.8\times 10^{-4}$ & $6.5\times 10^{-4}$ \\
LiteBIRD &  $1.5\times 10^{-3}$ &$1.3\times 10^{-3}$ & $1.5\times 10^{-3}$ \\
PICO &  $1.6\times 10^{-4}$ &$2.5\times 10^{-4}$ & $2.8\times 10^{-4}$ \\
\hline
\end{tabular}
\caption{Statistical error on $r$ without (second column) and with marginalization over the amplitude of a patchy-reionization power spectrum template corresponding to minimum masses of halos hosting ionizing sources of $M_{\text{min}}=10^9 \text{M}_\odot$ and $M_{\text{min}}=10^{11} \text{M}_\odot$. Results are shown for the future ground-based experiments, CMB-S4~\cite{2019arXiv190704473A} and Simons Observatory~\cite{SOforecast}, and the space missions LiteBIRD~\cite{Hazumi2019} and PICO~\cite{PICO2019}.} 

\label{tab:bias}
\end{table}

\section{Discussion}
We have investigated the imprint of patchy reionization on the cosmic
microwave background $B$-mode polarization. The $B$-mode power
from patchy reionization depends on the reionization history and
the morphology and clustering of the ionized regions.  Compared to
previous studies, we used realistic radiative transfer simulations of
reionization to estimate the $B$-mode signal.  These simulations are
calibrated to match the Thomson scattering optical depth measured by
\textit{Planck} and the mean transmission fraction of the Ly$\alpha$
forest.  With this calibration, these simulations have been found to match
successfully a variety of observations related to reionization,
such as the observed spatial fluctuations in the Ly$\alpha$ opacity of
the IGM, the ionization fraction inferred from the frequency of dark
pixels in Ly$\alpha$ forest spectra, quasar near zones, as well as the
Ly$\beta$ opacity of the IGM.  While these radiative transfer
simulations model reionization on scales up to $160h^{-1}\,\text{cMpc}$
(i.e., $236\,\text{cMpc}$), we extrapolate them up to scales of $1\,\text{cGpc}$ by using an
excursion-set-based approach.  In this model, reionization begins at
redshift $z\approx 15$, is halfway complete at $z=7$, and finishes at
$z=5.2$.

We find that the $B$-mode power at multipoles $l\approx 100$ due to scattering and screening of
CMB photons during reionization (scattering dominates on these scales) is more than one order of magnitude
lower than the primary $B$-mode from primordial gravitational waves with a tensor-to-scalar ratio 
$r = 10^{-3}$. The power from patchy reionization is comparable to the power of the primary signal for 
$r\sim 10^{-4}$ at these scales.  

Two factors play an important role in determining the amplitude of the
$B$-mode power spectrum induced by reionization: (i) the
ionization history; and (ii) the minimum mass of the halos hosting ionizing sources,
For a given minimum halo mass, early
reionization scenarios increase the induced
$B$-mode power spectrum, although such early reionization scenarios
appear to be ruled out by a variety of astrophysical data.  For a
given reionization history, the amplitude of the $B$-mode power
spectrum increases with the minimum mass of ionizing sources on the scales of interest.
At $\ell\sim 100$, even relatively extreme assumptions about the
ionization history or the minimum halo mass do not
increase the $B$-mode power spectrum beyond the primary signal for
$r\sim 10^{-3}$.

Near-future ground-based experiments, such as BICEP Array, Simons Observatory, and
CMB-S4~\cite{BICEPArray2018,SOforecast, CMBS42016}, aim to have the sensitivity to detect the primary $B$ modes around $\ell \sim 100$ for
  tensor-to-scalar ratios as low as $r \sim 10^{-3}$. For these experiments, we showed bias from not accounting for the $B$ modes induced by patchy reionization is small. Future space
missions such as LiteBIRD, PICO, and CMB-Bharat aim to target both the
$B$-mode signals generated by homogeneous reionization ($\ell \lesssim 20$) as well as recombination~\cite{2018SPIE10698E..1YS, 2018SPIE10698E..4FS,
  CMB-Bharat-Proposal}. Patchy reionization is a negligible contaminant to the former for detectable levels of primordial gravitational waves.

With sufficient sensitivity and aggressive delensing, future experiments may be able to reach 
$r<10^{-4}$ with measurements around the recombination peak. In this case, 
contamination from patchy reionization could become significant and might have to be
addressed in order to establish the nature of the measured signal. Marginalising over the amplitude of the patchy reionization power spectrum should be very effective at removing any bias, albeit with a modest increase in the statistical error on $r$, since the \emph{shape} of the patchy reionization power spectrum on large scales is very similar across plausible reionization models.
One could also consider partially removing the $B$ modes induced by patchy reionization in a process similar to delensing. 
This would additionally remove the cosmic variance of the patchy-reionization signal from the error on $r$.
The optical depth fluctuations can be reconstructed with quadratic estimators applied to the observed CMB fields~\cite{Dvorkin2009}. This reconstruction can be combined with the observed small-scale $E$-mode polarization to approximate the $B$ modes produced by patchy screening, which can be subtracted from the observed $B$ modes. The situation is more complicated for the $B$ modes produced by patchy scattering since one also needs the projection on the sky of the remote temperature quadrupole anisotropy through reionization. However, this can be estimated from the large-angle $E$-mode polarization (e.g., from a future space mission).

In conclusion, we do not expect the search for primordial $B$ modes
will be hindered by patchy reionization in the near future. For more
conventional reionization histories, all currently-planned ground-based CMB
experiments should be safe from significant contamination.

\acknowledgments{AR would like to thank the Kavli Institute for Cosmology Cambridge (KICC)
 for hospitality during the early phases of this project. 
 The authors would like to thank
  Marcelo Alvarez, Nicholas Battaglia, Tirthankar Roy Choudhury, Cora Dvorkin, Rishi
  Khatri, Ranjan Modak, David Spergel, and Matteo Viel for useful
  discussions.  The results presented in the paper relied on the
  Cambridge Service for Data Driven Discovery (CSD3) operated by the
  University of Cambridge (www.csd3.cam.ac.uk), provided by Dell EMC
  and Intel using Tier-2 funding from the Engineering and Physical
  Sciences Research Council (capital grant EP/P020259/1), and DiRAC
  funding from the Science and Technology Facilities Council
  (www.dirac.ac.uk).  This work further used the COSMA Data Centric
  system operated Durham University on behalf of the STFC DiRAC HPC
  Facility. This equipment was funded by a BIS National
  E-infrastructure capital grant ST/K00042X/1, DiRAC Operations grant
  ST/K003267/1 and Durham University. DiRAC is part of the National
  E-Infrastructure.  This work is partially supported by PRIN MIUR
  2015 grant `Cosmology and Fundamental Physics: illuminating the Dark
  Universe with Euclid', PRIN INAF 2014 grant `Probing the AGN/galaxy
  co-evolution through ultra-deep and ultra-high-resolution radio
  surveys', the MIUR grant `Finanziamento annuale individuale attivita
  base di ricerca' and the RADIOFOREGROUNDS grant (COMPET-05-2015,
  agreement number 687312) of the European Union Horizon 2020 research
  and innovation program. PDM acknowledges support from a Kavli
  Institute Senior Fellowship at the University of Cambridge and the
  Netherlands organization for scientific research (NWO) VIDI grant
  (dossier 639.042.730). MGH, GK and AR acknowledge support from ERC
  Advanced Grant 320596 `Emergence'.  AC and MH acknowledge support from the UK Science and Technology
  Facilities Council (grant numbers ST/N000927/1 and ST/S000623/1).
  While working on this paper, we
  learned about very similar work in preparation by Mukherjee et al.~\cite{2019arXiv190301994M}.  Although our simulation approach differs from theirs,
  we obtain similar conclusions. We have included comparison to their
  work in several places throughout this paper. }

\bibliography{mybib}
\bibliographystyle{JHEP}
\end{document}